\documentclass[12pt]{iopart}

\usepackage{iopams}
\usepackage{graphicx}		
\usepackage{pict2e}         
\usepackage{hyperref}		

\newcommand{\C}[2]{{#1 \choose #2}}
\renewcommand{\O}[1]{\mathcal{O}\left(#1\right)}
\newcommand{\arcsinh}{\mathrm{arcsinh}}
\newcommand{\Li}{\mathrm{Li}}
\newcommand{\I}{\mathrm{I}}
\renewcommand{\Re}{\mathrm{Re}}
\renewcommand{\Im}{\mathrm{Im}}
\newcommand{\deltaalpha}{\delta\!\alpha}
\newcommand{\deltabeta}{\delta\!\beta}

\newcommand{\affiliation}[1]{\address{#1}}
\renewcommand{\pacs}[1]{\noindent\textbf{PACS numbers:} #1}
\newcommand{\keywords}[1]{\noindent\textbf{Keywords:} #1}
\newcommand{\openone}{\mbox{{\small 1}$\!\!$1}}
\newcommand{\text}[1]{\mathrm{#1}}

\newcommand{\Up}[1]{\put #1{\line(1,1){10}}}
\newcommand{\Down}[1]{\put #1{\line(1,-1){10}}}
\newcommand{\UpThick}[1]{\put #1{\thicklines\line(1,1){10}}}
\newcommand{\DownThick}[1]{\put #1{\thicklines\line(1,-1){10}}}
\newcommand{\UpDotted}[1]{\multiput #1(1,1){10}{\put(0,0){$.$}}}
\newcommand{\DownDotted}[1]{\multiput #1(1,-1){10}{\put(0,-1){$.$}}}

\begin{document}

\title[Spectrum of periodic TASEP - bulk eigenvalues]{Spectrum of the totally asymmetric simple exclusion process on a periodic lattice - bulk eigenvalues}
\author{Sylvain Prolhac}
\affiliation{Laboratoire de Physique Th\'eorique, IRSAMC, UPS, Universit\'e de Toulouse, France\\Laboratoire de Physique Th\'eorique, UMR 5152, Toulouse, CNRS, France}
\date{\today}

\begin{abstract}
We consider the totally asymmetric simple exclusion process (TASEP) on a periodic one-dimensional lattice of $L$ sites. Using Bethe ansatz, we derive parametric formulas for the eigenvalues of its generator in the thermodynamic limit. This allows to study the curve delimiting the edge of the spectrum in the complex plane. A functional integration over the eigenstates leads to an expression for the density of eigenvalues in the bulk of the spectrum. The density vanishes with an exponent $2/5$ close to the eigenvalue $0$.\\

\\

\end{abstract}

\maketitle

\begin{section}{Introduction}
\label{section introduction}
Markov processes \cite{S70.1} form a class of mathematical models much studied in relation with non-equilibrium statistical physics. Their evolution in time is generated by an operator $M$, the Markov matrix, whose non-diagonal entries represent the rates at which the state of the system changes from a given microstate to another.\\\indent
For processes verifying the detailed balance condition, which forbids probability currents between the different microstates at equilibrium, the operator $M$ is real symmetric, up to a similarity transformation. It implies that its eigenvalues are real numbers. An example is the Ising model with \textit{e.g.} Glauber dynamics. Processes that do not satisfy detailed balance, on the other hand, generally have a complex spectrum. This is the case for the asymmetric simple exclusion process (ASEP) \cite{D98.1,S01.1,GM06.1,D07.1,S07.1,FS11.1,M11.1,CMZ11.1}, which consists of classical hard-core particles hopping between nearest neighbour sites of a lattice with a preferred direction.\\\indent
In one dimension, ASEP is known to be exactly solvable by means of Bethe ansatz. This has allowed exact calculations of the gap of the spectrum \cite{GS92.1,GS92.2,K95.1,GM04.1,GM05.1} and of the fluctuations of the current, both in the infinite line setting \cite{J00.1,PS02.1,BFPS07.1,TW09.1,SS10.1,ACQ11.1} and on a finite lattice with either periodic \cite{DL98.1,DA99.1,PM08.1,P08.1,PM09.1,P10.1} or open \cite{LM11.1,GLMV12.1,L13.1} boundary conditions. The exponents and scaling functions obtained in these articles are universal: they characterize not only driven diffusive systems \cite{KLS84.1,SZ98.1} far from equilibrium, to which ASEP belongs, but also interface growth models \cite{BS95.1,M98.1,HHZ95.1} and directed polymers in a random medium \cite{HHZ95.1,BD00.1,D10.1,CLDR10.1}. This forms the Kardar-Parisi-Zhang universality class \cite{KPZ86.1,KK10.1,SS10.4}.\\\indent
We focus in this article on the special case of ASEP with unidirectional hopping of the particles called totally asymmetric simple exclusion process (TASEP). We consider the system with periodic boundary conditions, for which the number of microstates is finite: the spectrum is then a discrete set of points in the complex plane. The aim of the present article is to obtain a large scale description of these points. We obtain explicit expressions for the curve delimiting the edge of the spectrum in the complex plane. This curve is singular near the eigenvalue $0$, with an imaginary part scaling as the real part to the power $5/3$. By a functional integration over the eigenstates, we also derive expressions for the density of eigenvalues. Near the eigenvalue $0$, the density vanishes with an exponent $2/5$.\\\indent
The paper is organized as follows: in section \ref{section TASEP}, we recall a few known things about TASEP and present our main results. In section \ref{section eigenvalues}, we derive parametric expressions for the eigenvalues of the Markov matrix of TASEP in the thermodynamic limit, and apply this to the curve delimiting the edge of the spectrum in section \ref{section envelope}. In section \ref{section density}, we study the density of eigenvalues. Finally in section \ref{section trace} we consider a generating function for the cumulants of the eigenvalues.
\end{section}

\begin{section}{Totally asymmetric exclusion process on a ring}
\label{section TASEP}
We consider in this article the totally asymmetric simple exclusion process (TASEP) with $N$ particles on a periodic one-dimensional lattice of $L$ sites (see fig.~\ref{fig TASEP}). A site is either empty or occupied by one particle. We call $\Omega$ the set of all microstates (or \textit{configurations}), which has cardinal $|\Omega|=\C{L}{N}$. A particle at site $i$ can hop to site $i+1$ if the latter is empty. The hopping rate for any particle is equal to $1$, \textit{i.e.} each particle allowed to move has a probability $\delta t$ in a small time interval $\delta t$. In the following, we call $\rho=N/L$ the density of particles.
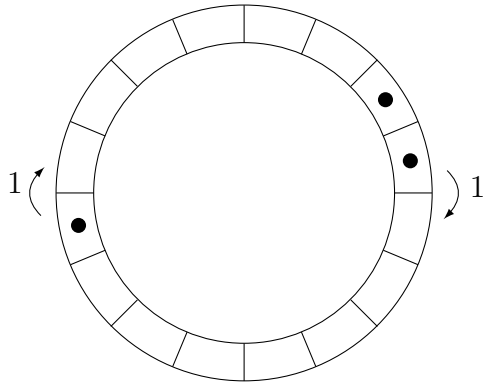
\begin{figure}
  \begin{center}
    \begin{picture}(70,50)
      \put(35,25){\circle{40}}
      \put(35,25){\circle{50}}
      \put(35,45){\line(0,1){5}}
      \put(42.6,43.48){\line(5,12){1.92}}
      \put(49.2,39.2){\line(1,1){3.54}}
      \put(53.48,32.6){\line(12,5){4.62}}
      \put(55,25){\line(1,0){5}}
      \put(53.48,17.4){\line(12,-5){4.62}}
      \put(49.2,10.8){\line(1,-1){3.54}}
      \put(42.6,6.52){\line(5,-12){1.92}}
      \put(35,5){\line(0,-1){5}}
      \put(27.4,6.52){\line(-5,-12){1.92}}
      \put(20.8,10.8){\line(-1,-1){3.54}}
      \put(16.52,17.4){\line(-12,-5){4.62}}
      \put(15,25){\line(-1,0){5}}
      \put(16.52,32.6){\line(-12,5){4.62}}
      \put(20.8,39.2){\line(-1,1){3.54}}
      \put(27.4,43.48){\line(-5,12){1.92}}
      \put(53.8,37.4){\circle*{2}}
      \put(57.1,29.3){\circle*{2}}
      \put(13,20.7){\circle*{2}}
      \qbezier(62,28)(65,25)(62,22)
      \put(62,22){\vector(-1,-1){0.5}}
      \qbezier(8,22)(5,25)(8,28)
      \put(8,28){\vector(1,1){0.5}}
      \put(65,24.5){$1$}
      \put(3.5,25){$1$}
    \end{picture}
    \caption{Totally asymmetric simple exclusion process with $N=3$ particles on a periodic one-dimensional lattice with $L=16$ sites. The particles hop to the nearest site in the clockwise direction with rate $1$ provided that site is empty.}
    \label{fig TASEP}
  \end{center}
\end{figure}

\begin{subsection}{Master equation}
\label{subsection master equation}
We call $P_{t}(\mathcal{C})$ the probability to observe the system in the microstate $\mathcal{C}$ at time $t$. The probabilities evolve in time by the \textit{master equation}
\begin{equation}
\label{master eq}
\frac{dP_{t}(\mathcal{C})}{dt}=\sum_{\mathcal{C}'\neq\mathcal{C}}\Big[w(\mathcal{C}\leftarrow\mathcal{C}')P_{t}(\mathcal{C}')-w(\mathcal{C}'\leftarrow\mathcal{C})P_{t}(\mathcal{C})\Big]\;.
\end{equation}
The rate $w(\mathcal{C}'\leftarrow\mathcal{C})$ is equal to $1$ if it is possible to go from configuration $\mathcal{C}$ to configuration $\mathcal{C}'$ by moving one particle to the next site, and $0$ otherwise.\\\indent
The master equation (\ref{master eq}) can be conveniently written as a matrix equation by defining the vector $|P_{t}\rangle=\sum_{\mathcal{C}\in\Omega}P_{t}(\mathcal{C})|\mathcal{C}\rangle$ of the configuration space $V$ with dimension $|\Omega|$, where $|\mathcal{C}\rangle$ is the canonical vector of $V$ corresponding to the configuration $\mathcal{C}$. Then, calling $M$ the matrix with non-diagonal entries $\langle\mathcal{C}|M|\mathcal{C}'\rangle=w(\mathcal{C}\leftarrow\mathcal{C}')$ and diagonal $\langle\mathcal{C}|M|\mathcal{C}\rangle=-\sum_{\mathcal{C}'\neq\mathcal{C}}w(\mathcal{C}'\leftarrow\mathcal{C})$, one has
\begin{equation}
\label{master eq matrix}
\frac{d}{dt}|P_{t}\rangle=M|P_{t}\rangle\;,
\end{equation}
which is formally solved in terms of the time evolution operator $\rme^{tM}$ as $|P_{t}\rangle=\rme^{tM}|P_{0}\rangle$.\\\indent
The graph of allowed transitions for TASEP presents an interesting cyclic structure with period $L$: let us consider the observable $X$ such that $X(C)$ is the sum of the positions of the particles (we take a fixed arbitrary site to be the origin of positions). We see that each time a particle hops to the next site, $X(C)$ increases of $1$ modulo $L$. It is then possible to split the configurations in $L$ sectors according to the value of $X$ modulo $L$. The only allowed transitions between configurations are then transitions from configurations of a sector $r$ to configurations of sector $r+1$ (modulo $L$), see fig.~\ref{fig graph dynamics TASEP}. We emphasize that this cyclic structure is not a consequence of the periodic boundary conditions. Indeed, a similar cyclic structure exists for ASEP on an open segment of $L$ sites connected to reservoirs of particles, with periodicity $L+1$ instead of $L$.
\begin{figure}
  \begin{center}
    \begin{picture}(130,115)
      \put(67.5,92.5){\circle*{2}}
      \put(72.5,92.5){\circle*{2}}
      \put(62.5,102.5){\circle*{2}}
      \put(77.5,102.5){\circle*{2}}
      \put(52.5,112.5){\circle*{2}}
      \put(57.5,112.5){\circle*{2}}
      \multiput(50,90)(0,10){3}{
        \multiput(0,0)(0,5){2}{\line(1,0){30}}
        \multiput(0,0)(5,0){7}{\line(0,1){5}}
      }
      \put(82.5,92.5){\vector(1,0){16.3}}
      \put(82.5,101.5){\vector(5,-2){18.3}}
      \put(82.5,103.5){\vector(9,-2){25.3}}
      \put(82.5,112.5){\vector(2,-1){26.3}}
      \put(112.5,77.5){\circle*{2}}
      \put(118.5,69.5){\circle*{2}}
      \put(111.5,95.5){\circle*{2}}
      \put(117.5,87.5){\circle*{2}}
      \multiput(100,90)(8,6){2}{
        \multiput(0,0)(4,3){2}{\line(3,-4){18}}
        \multiput(0,0)(3,-4){7}{\line(4,3){4}}
      }
      \put(119.8,65.2){\vector(-4,-9){7.2}}
      \put(121.8,66.2){\vector(-1,-10){2.3}}
      \put(127,71){\vector(-2,-9){6.2}}
      \put(128.8,72.2){\vector(0,-1){35.2}}
      \put(120.5,26.5){\circle*{2}}
      \put(123.5,30.5){\circle*{2}}
      \put(109.5,28.5){\circle*{2}}
      \put(118.5,40.5){\circle*{2}}
      \put(95.5,26.5){\circle*{2}}
      \put(98.5,30.5){\circle*{2}}
      \multiput(108,14)(-8,6){3}{
        \multiput(0,0)(4,-3){2}{\line(3,4){18}}
        \multiput(0,0)(3,4){7}{\line(4,-3){4}}
      }
      \put(93,23.5){\vector(-1,-1){10.3}}
      \put(100,18){\vector(-3,-1){17.3}}
      \put(102,17){\vector(-9,-5){20.3}}
      \put(108,12){\vector(-3,-1){25.3}}
      \put(62.5,2.5){\circle*{2}}
      \put(72.5,2.5){\circle*{2}}
      \put(57.5,12.5){\circle*{2}}
      \put(77.5,12.5){\circle*{2}}
      \multiput(50,0)(0,10){2}{
        \multiput(0,0)(0,5){2}{\line(1,0){30}}
        \multiput(0,0)(5,0){7}{\line(0,1){5}}
      }
      \put(47,13.5){\vector(-1,1){10.3}}
      \put(47,12.33333){\vector(-3,1){17.3}}
      \put(48,5.88888){\vector(-9,5){20.3}}
      \put(47,3.66666){\vector(-3,1){25.3}}
      \put(12.5,22.5){\circle*{2}}
      \put(9.5,26.5){\circle*{2}}
      \put(14.5,36.5){\circle*{2}}
      \put(23.5,24.5){\circle*{2}}
      \put(19.5,46.5){\circle*{2}}
      \put(34.5,26.5){\circle*{2}}
      \multiput(22,14)(8,6){3}{
        \multiput(0,0)(-4,-3){2}{\line(-3,4){18}}
        \multiput(0,0)(-3,4){7}{\line(-4,-3){4}}
      }
      \put(17.2,49.45){\vector(-4,9){7.2}}
      \put(10.4,44.2){\vector(-1,10){2.3}}
      \put(9,44){\vector(-2,9){6.2}}
      \put(1.2,37.3){\vector(0,1){35.3}}
      \put(14.5,73.5){\circle*{2}}
      \put(26.5,89.5){\circle*{2}}
      \put(9.5,83.5){\circle*{2}}
      \put(15.5,91.5){\circle*{2}}
      \multiput(30,90)(-8,6){2}{
        \multiput(0,0)(-4,3){2}{\line(-3,-4){18}}
        \multiput(0,0)(-3,-4){7}{\line(-4,3){4}}
      }
      \put(22.5,97.5){\vector(5,-1){25.3}}
      \put(29.5,91.5){\vector(9,5){18.3}}
      \put(22.5,99.33333){\vector(6,1){25.3}}
      \put(28.5,93.5){\vector(1,1){19.3}}
    \end{picture}
    \caption{Graph of all allowed transitions between the configurations of TASEP with $N=2$ particles on a periodic lattice of $L=6$ sites.}
    \label{fig graph dynamics TASEP}
  \end{center}
\end{figure}
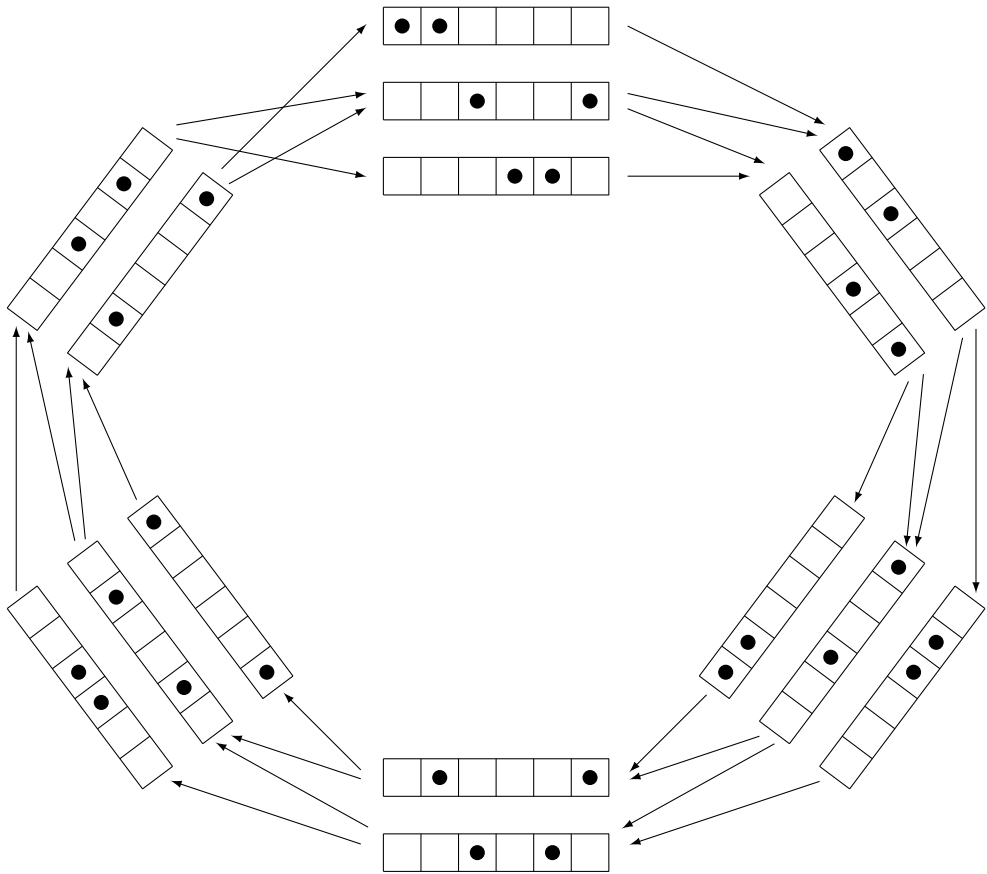
\end{subsection}

\begin{subsection}{Current fluctuations}
\label{subsection current}
Equal time observables, such as the density profile of particles in the system or the number of clusters of consecutive particles can be extracted directly from the knowledge of $|P_{t}\rangle$. Other observables, however, require the $P_{t}(\mathcal{C})$ for several values of the time. A much studied example is the current of particles and especially the fluctuations around its mean value.\\\indent
We define the observable $Y_{t}$, which counts the total displacement of particles between time $0$ and time $t$. Starting with $Y_{0}=0$, it is then updated by $Y_{t}\to Y_{t}+1$ each time a particle hops anywhere in the system.\\\indent
The joint probability $P_{t}(\mathcal{C},Y)$ to observe the system in the configuration $\mathcal{C}$ with $Y_{t}=Y$ obeys the master equation
\begin{equation}
\label{master eq Y}
\frac{dP_{t}(\mathcal{C},Y)}{dt}=\sum_{\mathcal{C}'\neq\mathcal{C}}\Big[w(\mathcal{C}\leftarrow\mathcal{C}')P_{t}(\mathcal{C}',Y-1)-w(\mathcal{C}'\leftarrow\mathcal{C})P_{t}(\mathcal{C},Y)\Big]\;.
\end{equation}
It is convenient to introduce the quantity $F_{t}(\mathcal{C})=\sum_{Y=-\infty}^{\infty}\rme^{\gamma Y}P_{t}(\mathcal{C},Y)$. It verifies the deformed master equation \cite{DL98.1}
\begin{equation}
\label{master eq gamma}
\frac{dF_{t}(\mathcal{C})}{dt}=\sum_{\mathcal{C}'\neq\mathcal{C}}\Big[\rme^{\gamma}w(\mathcal{C}\leftarrow\mathcal{C}')F_{t}(\mathcal{C}')-w(\mathcal{C}'\leftarrow\mathcal{C})F_{t}(\mathcal{C})\Big]\;.
\end{equation}
Introducing the vector $|F_{t}\rangle=\sum_{\mathcal{C}\in\Omega}F_{t}(\mathcal{C})|\mathcal{C}\rangle$ and a deformation $M(\gamma)$ of the Markov matrix, one has
\begin{equation}
\label{master eq matrix}
\frac{d}{dt}|F_{t}\rangle=M(\gamma)|F_{t}\rangle\;.
\end{equation}
For $\gamma=0$, $F_{t}$ reduces to $P_{t}$ and $M(0)=M$. In the following, we will be interested in the spectrum of the operators $M$ and $M(\gamma)$.
\end{subsection}

\begin{subsection}{Mapping to a height model and continuous spectrum}
It is well known that TASEP can be mapped to a model of growing interface \cite{HHZ95.1}: for each occupied site $i$ of the system, we draw a portion of interface decreasing from  height $h_{i}$ to $h_{i+1}=h_{i}-(1-\rho)$, and for each empty site $i$, we draw a portion of interface increasing from height $h_{i}$ to $h_{i+1}=h_{i}+\rho$. The interface obtained is continuous and periodic, see fig.~\ref{fig growth model}. The dynamics of TASEP then implies that parallelograms (squares at half-filling $N=L/2$) deposit on local minima of the interface with rate 1.\\\indent
Unlike TASEP, the set of microstates of the growth model is not finite since the total height is not bounded: there always exists a local minimum from which the interface can grow. It is possible to identify any microstate of the growth model by the corresponding configuration of the exclusion process and the total current $Y$ defined in section \ref{subsection current}. Then, (\ref{master eq Y}) can be interpreted as the master equation for the growth model, to which is associated the infinite dimensional Markov matrix
\begin{equation}
\mathcal{M}=M^{(+)}\otimes S+M^{(0)}\otimes\openone\;,
\end{equation}
where $M^{(0)}$ and $M^{(+)}$ are respectively the diagonal and non-diagonal part of $M$, and $S$ is the translation operator in $Y$ space, $S=\sum_{Y=-\infty}^{\infty}|Y\rangle\langle Y-1|$.\\\indent
We would like to diagonalize $\mathcal{M}$ in order to study its spectrum. We note that $\mathcal{M}$ commutes with $S$.
This implies that the eigenvectors $|\Psi\rangle$ of $\mathcal{M}$ must be of the form
\begin{equation}
|\Psi\rangle=|\mathcal{\psi}\rangle\otimes|\phi(\theta)\rangle\;,
\end{equation}
where $|\psi\rangle$ is a vector in configuration space and
\begin{equation}
|\phi(\theta)\rangle=\frac{1}{\sqrt{2\pi}}\sum_{Y=-\infty}^{\infty}\rme^{-\rmi\,\theta\,Y}|Y\rangle\;
\end{equation}
the right eigenvector of $S$ with eigenvalue $\rme^{\rmi\,\theta}$. The eigenvalue equation for $|\Psi\rangle$ can be written
\begin{equation}
E|\Psi\rangle=[(\rme^{\rmi\,\theta}M^{(+)}+M^{(0)})|\psi\rangle]\otimes|\phi(\theta)\rangle=[M(\rmi\,\theta)|\psi\rangle]\otimes|\phi(\theta)\rangle\;,
\end{equation}
where $M(\rmi\,\theta)$ is the deformation of the Markov matrix introduced in section \ref{subsection current} and $|\psi\rangle$ and $E$ an eigenvector and an eigenvalue of $M(\rmi\,\theta)$. We finally find that the spectrum of $\mathcal{M}$ is the reunion of the spectra of the $M(\gamma)$ with $|\rme^{\gamma}|=1$. This spectrum is represented in fig.~\ref{fig continuous spectrum} for $L=8$, $N=4$.\\\indent
The spectrum of $\mathcal{M}$ can also be constructed from the finite spectra obtained by counting the total current $Y$ modulo $KL$ with $K$ a positive integer, and taking the limit $K\to\infty$. The case $K=1$ is the usual TASEP because of the cyclic structure of the graph of allowed transitions discussed at the end of section \ref{subsection master equation}. The case $K=N$ corresponds to TASEP with distinguishable particles, restricted to a subspace with a given cyclic order of the particles since the particles cannot overtake each other.\\\indent
We call $M^{(K)}(\gamma)$ the corresponding deformed Markov matrix. The $K|\Omega|$ configurations arrange themselves in $KL$ sectors according to the value of $Y$. Calling $P_{r}$ the projector on the $r$-th sector and $M_{r,r+1}^{(K)}(\gamma)=\rme^{\gamma}P_{r+1}M^{(K)}P_{r}+P_{r}M^{(K)}P_{r}$, one has
\begin{equation}
M^{(K)}(\gamma)=\sum_{r=1}^{KL}M_{r,r+1}^{(K)}(\gamma)\;.
\end{equation}
Introducing $U=\sum_{r=1}^{KL}\rme^{r\gamma}P_{r}$, one finds
\begin{equation}
U^{-1}M^{(K)}(\gamma)U=\sum_{r=1}^{KL-1}M_{r,r+1}^{(K)}+\rme^{KL\gamma}P_{1}M_{KL,1}^{(K)}P_{KL}+P_{KL}M_{KL,1}^{(K)}P_{KL}\;,
\end{equation}
with $M_{r,r+1}^{(K)}=M_{r,r+1}^{(K)}(0)$. This implies that the spectrum of $M^{(K)}(\gamma)$ is invariant under the transformation $\gamma\to\gamma+2\rmi\pi/(KL)$. Starting from a given eigenvalue of $M^{(K)}(\gamma)$ and following the eigenvalue during the continuous change from $\gamma$ to $\gamma+2\rmi\pi/(KL)$, one does not in general come back to the initial eigenvalue: we observe by numerical diagonalization that one goes from one eigenvalue to the next one anticlockwise on the same continuous curve in fig.~\ref{fig continuous spectrum}.
\begin{figure}
  \begin{center}
    \begin{picture}(140,70)
      \put(15,3){\circle*{2}}
      \put(35,3){\circle*{2}}
      \put(45,3){\circle*{2}}
      \put(85,3){\circle*{2}}
      \put(95,3){\circle*{2}}
      \put(105,3){\circle*{2}}
      \qbezier(45,5)(50,10)(55,5)
      \put(55,5){\vector(1,-1){0.2}}
      \put(50,10){$1$}
      \put(10,0){\line(1,0){120}}
      \put(10,0){\line(0,1){5}}
      \put(20,0){\line(0,1){5}}
      \put(30,0){\line(0,1){5}}
      \put(40,0){\line(0,1){5}}
      \put(50,0){\line(0,1){5}}
      \put(60,0){\line(0,1){5}}
      \put(70,0){\line(0,1){5}}
      \put(80,0){\line(0,1){5}}
      \put(90,0){\line(0,1){5}}
      \put(100,0){\line(0,1){5}}
      \put(110,0){\line(0,1){5}}
      \put(120,0){\line(0,1){5}}
      \put(130,0){\line(0,1){5}}
      \UpDotted{(10,20)}
      \DownDotted{(20,30)}
      \UpDotted{(30,20)}
      \DownThick{(40,30)}
      \UpThick{(50,20)}
      \DownDotted{(60,30)}
      \UpDotted{(70,20)}
      \DownDotted{(80,30)}
      \UpDotted{(90,20)}
      \DownThick{(100,30)}
      \UpThick{(110,20)}
      \DownDotted{(120,30)}
      \DownThick{(10,40)}
      \UpThick{(20,30)}
      \DownThick{(30,40)}
      \UpThick{(60,30)}
      \DownThick{(70,40)}
      \UpThick{(80,30)}
      \DownThick{(90,40)}
      \UpThick{(120,30)}
      \put(50,37){\vector(0,-1){14}}
      \put(51,30){$1$}
      \Down{(40,50)}
      \Up{(50,40)}
      \Up{(40,50)}
      \Down{(50,60)}
    \end{picture}
    \caption{Mapping between periodic TASEP and an interface growth model at half-filling.}
    \label{fig growth model}
  \end{center}
\end{figure}
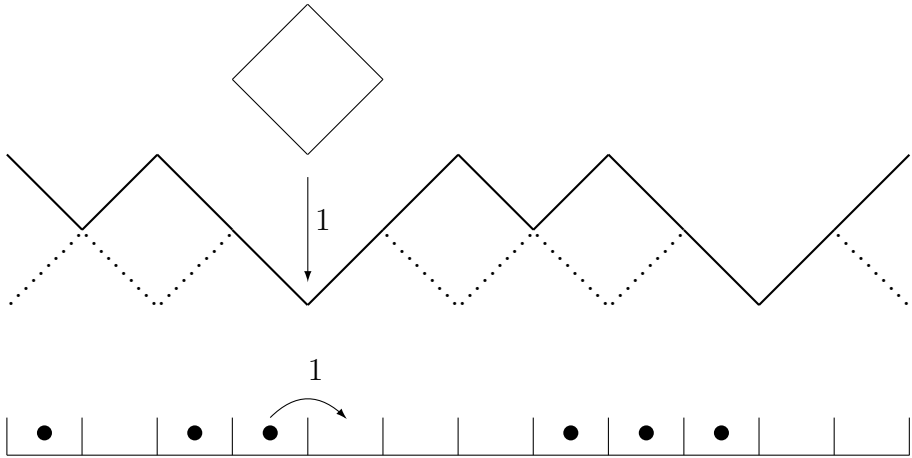
\begin{figure}
  \begin{center}
    \begin{tabular}{c}\includegraphics[width=70mm]{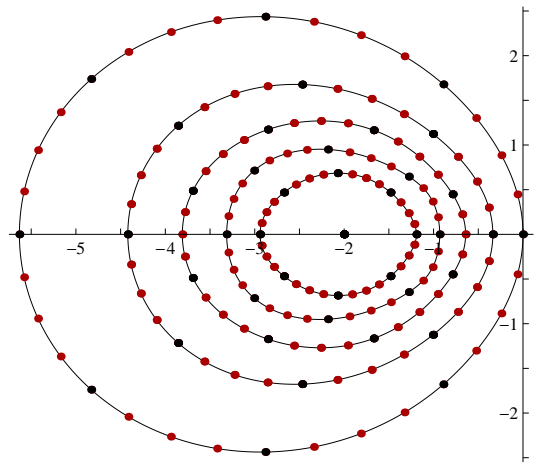}\end{tabular}
    \caption{Spectrum of the Markov matrix of TASEP with $4$ particles on $8$ sites. The black dots are the eigenvalues for undistinguishable particles, while the black $+$ red (gray in printed version) dots are the eigenvalues for distinguishable particles. The curves are the eigenvalues of the corresponding interface growth model.}
    \label{fig continuous spectrum}
  \end{center}
\end{figure}
\end{subsection}

\begin{subsection}{Large scale description of the spectrum}
Typical eigenvalues of the Markov matrix $M$ scale proportionally with $L$. Dividing all the eigenvalues by $L$, the rescaled spectrum fills a region of the complex plane in the thermodynamic limit. We call $e(\mu)$ a parametrization of the curve at the edge of this region. An exact representation of this curve, (\ref{e[d]}), (\ref{d'[d]}), is obtained in section \ref{section envelope}. We observe that $e(\mu)$ is singular near $e=0$, with the scaling
\begin{equation}
\label{Ree[Ime]}
\Re\,e\simeq-\frac{2^{2/3}3^{7/6}\pi^{2/3}}{10}|\Im\,e|^{5/3}\;.
\end{equation}
The density of eigenvalues $D(e)$ in the rescaled spectrum is also studied in section \ref{section density}. It grows for large $L$ as
\begin{equation}
\label{D[s]}
D(e)\sim\rme^{L s(e)}\;.
\end{equation}
The quantity $s(e)$ is computed exactly. For eigenvalues close to $0$, but far from the edge of the spectrum (\textit{i.e.} with $|\Im\,e|^{5/3}\ll|\Re\,e|$), one finds in particular
\begin{equation}
\label{s(e)}
\displaystyle s(e)\simeq\xi(-\Re\,e)^{2/5}
\quad
\text{with}
\quad
\xi\simeq1.58600
\;.
\end{equation}
The exponent $2/5$ is different from the exponent $1/3$ obtained in \ref{appendix free particles} for undistinguishable non-interacting particles hopping unidirectionally on a periodic one-dimensional lattice.
\end{subsection}

\begin{subsection}{Trace of the time evolution operator and cumulants of the eigenvalues}
\label{subsection Q}
We consider the quantity
\begin{equation}
\label{Q(t)}
Q(t)=\frac{1}{|\Omega|}\tr\rme^{tM}=\frac{1}{|\Omega|}\sum_{\mathcal{C}\in\Omega}\langle\mathcal{C}|\rme^{tM}|\mathcal{C}\rangle\;.
\end{equation}
Here $\langle\mathcal{C}|\rme^{tM}|\mathcal{C}\rangle$ is the probability that the system is in the microstate $\mathcal{C}$ at time $t$ conditioned on the fact it was already in the microstate $\mathcal{C}$ at time $0$. Since all configurations are equally probable in the stationary state of periodic TASEP \cite{D98.1}, $Q(t)$ is simply the stationary probability that the system is in the same microstate at both times $0$ and $t$.

\begin{subsubsection}{Perturbative expansion}
\hspace{0mm}\\\indent
We consider more generally (\ref{Q(t)}) with $M$ replaced by the deformation $M(\gamma)$, and define
\begin{equation}
\label{f[M]}
f(t)=\frac{1}{L}\log\tr\rme^{tM(\gamma)}\;.
\end{equation}
The quantity $f(t)$ can be seen as the generating function of the cumulants of the eigenvalues of $M(\gamma)$ (or more precisely the cumulants of the uniform probability distribution on the set of the eigenvalues). The moments $\mu_{k}=|\Omega|^{-1}L^{-k}\tr M(\gamma)^{k}$ and the cumulants $c_{k}=L^{1-k}f^{(k)}(0)$ of the eigenvalues are indeed related from (\ref{f[M]}) by
\begin{equation}
\log\Big(1+\sum_{k=1}^{\infty}\frac{\mu_{k}t^{k}}{k!}\Big)=\sum_{k=1}^{\infty}\frac{c_{k}t^{k}}{k!}\;.
\end{equation}
The first cumulants are $c_{1}=\mu_{1}=f'(0)$, $c_{2}=\mu_{2}-\mu_{1}^{2}=f''(0)/L$ and $c_{3}=\mu_{3}-3\mu_{1}\mu_{2}+2\mu_{1}^{3}=f'''(0)/L^{2}$\\\indent
The moments and cumulants of the eigenvalues are independent of the deformation $\gamma$. Indeed, one can write
\begin{equation}
\label{tr(Mk)}
\tr M(\gamma)^{k}=\sum_{\mathcal{C}_{1},\ldots,\mathcal{C}_{k}}\rme^{\gamma\sum_{j=1}^{k}\openone_{\{\mathcal{C}_{j}\neq\mathcal{C}_{j+1}\}}}\langle\mathcal{C}_{1}|M|\mathcal{C}_{2}\rangle\langle\mathcal{C}_{2}|M|\mathcal{C}_{3}\rangle\ldots\langle\mathcal{C}_{k}|M|\mathcal{C}_{1}\rangle\;.
\end{equation}
Because of the cyclic structure of allowed transitions explained at the end of section \ref{subsection master equation}, only $k$-tuples of configurations such that $\sum_{j=1}^{k}\openone_{\{\mathcal{C}_{j}\neq\mathcal{C}_{j+1}\}}$ is divisible by L contribute to (\ref{tr(Mk)}). For $k<L$, it implies that $\tr M(\gamma)^{k}$ cannot depend on $\gamma$. One has then
\begin{equation}
f(t)=\Big(\frac{1}{L}\log\tr\rme^{tM}+\mathcal{O}(t^{L})\Big)\;.
\end{equation}
It is possible to calculate directly the coefficients of the expansion near $t=0$ of $f(t)$ by considering the case $\gamma\to-\infty$, for which the matrix $M(\gamma)$ becomes diagonal in configuration basis: $\langle\mathcal{C}|M(\gamma)|\mathcal{C}\rangle$ is equal to minus the number $m(\mathcal{C})$ of clusters of consecutive particles in the system. It implies
\begin{equation}
\label{f[M] gamma=-infinity}
f(t)=\frac{1}{L}\sum_{\mathcal{C}\in\Omega}\rme^{-t m(\mathcal{C})}=\frac{1}{L}\sum_{m=1}^{N}|\Omega(m)|\rme^{-tm}\;.
\end{equation}
The total number $|\Omega(m)|$ of configurations with $m$ clusters can be calculated in the following way: a configuration with $m$ clusters for which the last site is occupied can be described as $a(0)\geq0$ particles followed by $b(1)>0$ empty sites, $a(1)>0$ particles, \ldots, $b(m)>0$ empty sites and $a(m)>0$ particles. The total number of such configurations is $A_{m+1}(N+1)A_{m}(L-N)$ with
\begin{equation}
A_{m}(r)=\sum_{b(1),\ldots,b(m)=1}^{\infty}\openone_{\{b(1)+\ldots+b(m)=r\}}=\oint\frac{\rmd z}{2\rmi\pi}\,\frac{z^{m-r-1}}{(1-z)^{m}}=\C{r-1}{m-1}\;.
\end{equation}
From particle-hole symmetry, the number of configurations with $m$ clusters for which the last site is empty is $A_{m+1}(L-N+1)A_{m}(N)$
This implies
\begin{equation}
|\Omega(m)|=\frac{mL}{N(L-N)}\,\C{N}{m}\C{L-N}{m}\;.
\end{equation}
A saddle point approximation of the sum over $m$ in (\ref{f[M] gamma=-infinity}) finally gives
\begin{eqnarray}
\label{f(t)}
&& f(t)=\rho\log\Big(\frac{(1-2\rho)+2\rho\rme^{t}+\sqrt{1+4\rho(1-\rho)(\rme^{t}-1)}}{2\rho\rme^{t}}\Big)\nonumber\\
&& +(1-\rho)\log\Big(\frac{-(1-2\rho)+2(1-\rho)\rme^{t}+\sqrt{1+4\rho(1-\rho)(\rme^{t}-1)}}{2(1-\rho)\rme^{t}}\Big)\;,
\end{eqnarray}
which simplifies at half-filling to
\begin{equation}
\label{f(t) rho=1/2}
f(t)=\log(1+\rme^{-t/2})\;.
\end{equation}
The expressions (\ref{f(t)}) and (\ref{f(t) rho=1/2}) must be understood as an equality between Taylor series.\\\indent
In section \ref{section trace}, we consider again the quantity $f(t)$ as a testing ground for the formulas derived from Bethe ansatz in section \ref{section eigenvalues} for the eigenvalues of $M$ in the thermodynamic limit. We write the summation over the eigenvalues as an integral over a function $\eta$ that index the eigenstates. After a saddle point calculation in the functional integral, we recover (\ref{f(t)}).
\end{subsubsection}

\begin{subsubsection}{Finite \texorpdfstring{$t$}{t}}
\hspace{0mm}\\\indent
The total number of particles hopping during a finite time $t$ is roughly proportional to the average number of clusters of consecutive particles in the system. For typical configurations, this number scales proportionally with the system size in the thermodynamic limit at a finite density of particles. Because of the cyclic structure with period $L$ in the graph of allowed transitions described at the end of section \ref{subsection master equation}, the quantity $f(t)$ should have oscillations for finite times. The same reasoning also works for undistinguishable non-interacting particles. For distinguishable particles, on the other hand, a similar argument shows that $f(t)$ should show oscillations on the scale $t\sim L$, since the cyclic structure of the graph of allowed transitions has then period $NL$: all the particles need to come back to their initial state.\\\indent
The oscillations of $f(t)$ are observed for TASEP from numerical diagonalization, see fig.~\ref{fig f}. For non-interacting particles, they are confirmed by a direct calculation in \ref{appendix free particles}, see fig.~\ref{fig f free}. In both cases, we observe that the oscillations of $f(t)$ are not smooth: the function $f(t)$ is defined piecewise. There exists in particular a time $t(\gamma)$ such that $f(t)$ is analytic (and independent of $\gamma$) for $t$ between $0$ and $t(\gamma)$. We find $t(\gamma)\simeq1.7085\,\rme^{-\gamma}$ for undistinguishable free particles. For TASEP at half-filling, fig.~\ref{fig f} seems to indicate that $t(0)$ is slightly larger than $1$.
\begin{figure}
  \begin{center}
    \begin{tabular}{c}\includegraphics[width=100mm]{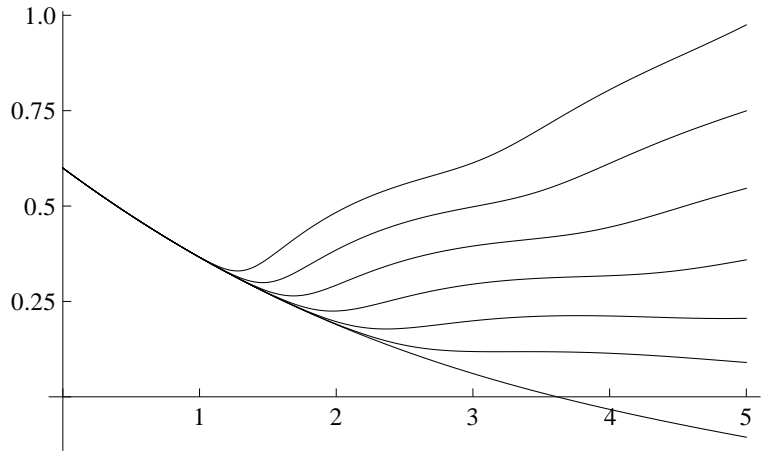}\end{tabular}
  \end{center}
  \caption{Plots of $L^{-1}\log\tr\rme^{tM(\gamma)}$ as a function of $t$ for a system at half-filling with $L=18$ sites, obtained from numerical diagonalization of the (deformed) Markov matrix. The different curves correspond to $\gamma=-\infty,0,0.1,0.2,0.3,0.4,0.5$, from bottom to top.}
  \label{fig f}
\end{figure}
\end{subsubsection}
\end{subsection}
\end{section}

\begin{section}{Parametric formulas for the eigenvalues}
\label{section eigenvalues}
In this section, we derive an exact parametric expression, (\ref{e[k]}), (\ref{b[k]}), for all the eigenvalues of the Markov matrix of TASEP.

\begin{subsection}{Bethe ansatz}
The deformed Markov matrix $M(\gamma)$ of TASEP is equivalent by similarity transformation to minus the Hamiltonian of a ferromagnetic XXZ spin chain with anisotropy $\Delta=\infty$ and twisted boundary conditions \cite{GM06.1}. The integrability of TASEP is a consequence of this. The eigenfunctions of $M(\gamma)$ are given by the Bethe ansatz as \cite{D98.1,S01.1,GM06.1}
\begin{equation}
\label{eigenvector}
\psi(x_{1},\ldots,x_{N})=\det\Big[\Big(\frac{y_{k}}{y_{j}}\Big)^{N-j}\rme^{\gamma x_{j}}(1-y_{k})^{x_{j}}\Big]_{j,k=1,\ldots,N}\;,
\end{equation}
provided the quantities $y_{j}$, called Bethe roots, verify the Bethe equations
\begin{equation}
\label{Bethe eq}
\frac{(1-y_{j})^{L}}{y_{j}^{N}}=(-1)^{N-1}\rme^{-L\gamma}\prod_{k=1}^{N}\frac{1}{y_{k}}\;.
\end{equation}
The Bethe equations have many different solutions, corresponding to the various eigenstates of $M(\gamma)$. The eigenvalue of $M(\gamma)$ corresponding to (\ref{eigenvector}) is
\begin{equation}
\label{E[y]}
E=\sum_{j=1}^{N}\frac{y_{j}}{1-y_{j}}\;.
\end{equation}
\textit{Remark}: the Bethe ansatz is usually written in terms of the variables $z_{j}=\rme^{\rmi q_{j}}=\rme^{\gamma}(1-y_{j})$ instead of the $y_{j}$'s. The eigenvectors are then linear combinations of plane waves with momenta $q_{j}$.\\
\textit{Remark 2}: proving the completeness of the Bethe ansatz for periodic TASEP is still an open problem, although one observes numerically for small systems that it does give all the eigenstates. The main difficulties consist in proving that the Bethe equations (\ref{Bethe eq}) have $|\Omega|$ solutions, and that all the eigenvectors generated form a basis of the configuration space of the model. Alternatively, the completeness would follow from a direct proof of the resolution of the identity, $\openone=\sum_{k=1}^{|\Omega|}|\psi_{k}\rangle\langle\psi_{k}|$. For periodic TASEP in a discrete time setting with parallel update, such a proof was given by Povolotsky and Priezzhev in \cite{PP07.1}.
\end{subsection}

\begin{subsection}{Parametric solution of the Bethe equations}
The Bethe equations (\ref{Bethe eq}) of TASEP have the particularity that the rhs does not depend specifically on $y_{j}$, but is instead a symmetric function of all the $y_{k}$'s. We write this rhs $(-1)^{N-1}/B^{L}$. This allows to solve the Bethe equations in a parametric way, by first solving for each $y_{j}$ the polynomial equation
\begin{equation}
\label{polynomial[y,B]=0}
(1-y_{j})^{L}/y_{j}^{N}=(-1)^{N-1}/B^{L}\;
\end{equation}
as a function of $B$, and then solving a self-consistency equation for $B$. This "decoupling property" was already used in \cite{GS92.1,GS92.2,GM04.1,GM05.1} for the calculation of the gap, and in \cite{DL98.1,DA99.1} for the calculation of the eigenvalue of $M(\gamma)$ with largest real part. The same property is also true for periodic TASEP in a discrete time setting with parallel update \cite{PP07.1}.\\\indent
Taking the power $1/L$ of the Bethe equations, there must exist numbers $k_{j}$, $j=1,\ldots,N$, integers if $N$ is odd, half-integers if $N$ is even, such that
\begin{equation}
\label{BE 1/L}
\frac{1-y_{j}}{y_{j}^{\rho}}=\frac{\rme^{2\rmi\pi k_{j}/L}}{B}\;,
\end{equation}
with $B$ a solution of
\begin{equation}
\label{B[y]}
\log B=\gamma+\frac{1}{L}\sum_{j=1}^{N}\log y_{j}\;.
\end{equation}
We also define
\begin{equation}
\label{l[rho]}
\ell=-\rho\log\rho-(1-\rho)\log(1-\rho)\;.
\end{equation}
The Bethe equations (\ref{BE 1/L}) involve a function $g$, defined as
\begin{eqnarray}
\label{g}
&& g:\mathbb{C}\backslash\mathbb{R^{-}}\to\mathbb{C}\backslash(\rme^{\rmi\pi\rho}[\rme^{\ell},\infty[\;\cup\;\rme^{-\rmi\pi\rho}[\rme^{\ell},\infty[)\nonumber\\
&&\hspace{14mm} y\mapsto\frac{1-y}{y^{\rho}}\;.
\end{eqnarray}
It turns out that this function is a \textit{bijection} for $0<\rho<1$, see \ref{appendix g}. This is a key point, as it allows to formally solve the Bethe equations as
\begin{equation}
\label{y[g,k]}
y_{j}=g^{-1}(\rme^{2\rmi\pi k_{j}/L}/B)\;,
\end{equation}
Then, the corresponding eigenvalue can be computed from (\ref{E[y]}), and $B$ is fixed by solving (\ref{B[y]}). We assumed that $\rme^{2\rmi\pi k_{j}/L}/B$ does not belong to the cut of $g^{-1}$. If this is not the case for some eigenstate, a continuity argument in the parameter $\gamma$ should still allow to use (\ref{y[g,k]}).\\
\textit{Remark}: there are exactly $|\Omega|$ ways to choose the $k_{j}$'s with the constraint $0<k_{1}<\ldots<k_{N}\leq L$. We observe numerically on small systems that all $|\Omega|$ eigenstates are recovered with this constraint. A similar argument was given in \cite{GM04.1}, using instead a rewriting of the Bethe equations as a polynomial equation of degree $L$ for the $z_{j}$'s: the number of ways to choose $N$ roots from this polynomial equation is also equal to $|\Omega|$.
\end{subsection}

\begin{subsection}{Functions \texorpdfstring{$\varphi$}{phi} and \texorpdfstring{$\psi$}{psi}}
We define the rescaled eigenvalue $e=E/L$ and the parameter $b=\log B$. We introduce the functions
\begin{equation}
\label{phipsi[g]}
\varphi(z)=\frac{g^{-1}(z)}{1-g^{-1}(z)}
\quad\text{and}\quad
\psi(z)=\log g^{-1}(z)\;.
\end{equation}
In terms of $\varphi$ and $\psi$, the parametric expression (\ref{E[y]}), (\ref{B[y]}) rewrites as
\begin{equation}
\label{e[k]}
\displaystyle e=\frac{1}{L}\sum_{j=1}^{n}\varphi\Big(\rme^{\frac{2\rmi\pi k_{j}}{L}-b}\Big)\;
\end{equation}
\begin{equation}
\label{b[k]}
\displaystyle b=\gamma+\frac{1}{L}\sum_{j=1}^{n}\psi\Big(\rme^{\frac{2\rmi\pi k_{j}}{L}-b}\Big)\;.
\end{equation}
From (\ref{phipsi[g]}) and (\ref{g}), the functions $\varphi$ and $\psi$ verify the relations
\begin{eqnarray}
\label{eq phi}
&& \log z+\rho\log\varphi(z)+(1-\rho)\log(1+\varphi(z))=0\\
\label{eq psi}
&& \rme^{\psi(z)}+z\,\rme^{\rho\,\psi(z)}=1\;.
\end{eqnarray}
They are also related by the two equations
\begin{eqnarray}
\label{relation phi psi 1}
&& \varphi(z)=\frac{\rme^{\psi(z)}}{1-\rme^{\psi(z)}}\\
\label{relation phi psi 2}
&& z\,\varphi(z)=\rme^{(1-\rho)\psi(z)}\;.
\end{eqnarray}
The expansion near $z=0$ of $\varphi(z)$ and $\psi(z)$ at arbitrary filling can be computed explicitly from (\ref{phipsi[g]}) and the observation \cite{DL98.1} that for any meromorphic function $h$, $h(g^{-1}(z))$ can be written as a contour integral. One has
\begin{equation}
\label{h(g-1)}
h(g^{-1}(z))=\oint_{g(\Gamma)}\frac{\rmd w}{2\rmi\pi}\frac{h(g^{-1}(w))}{w-z}=\oint_{\Gamma}\frac{\rmd y}{2\rmi\pi}\frac{g'(y)h(y)}{g(y)-z}\;,
\end{equation}
where the contour $\Gamma$ encloses $g^{-1}(z)$ but none of the poles of $h$, and does not cross the cut $\mathbb{R^{-}}$ of $g$. We will need to expand $h(g^{-1}(z))$ for small $z$. In the previous expression, it is possible to expand the integrand near $z=0$ as long as $|g(y)|>|z|$ for all $y\in\Gamma$. As shown in fig.~\ref{fig curves g}, it is possible to find such a contour $\Gamma$ only if $|z|<\rme^{\ell}$, otherwise the contour would have to go through the cut.\\\indent
We use (\ref{h(g-1)}) for $h(y)=\log y$ and expand for small $z$ inside the integral. After computing the residues at $y=1$ (which is always inside the contour, see fig.~\ref{fig curves g}), we obtain
\begin{equation}
\label{psi(z)}
\psi(z)=\sum_{r=1}^{\infty}\C{\rho\,r}{r}\frac{(-1)^{r}z^{r}}{\rho\,r}\;.
\end{equation}
Using (\ref{h(g-1)}) again for $h(y)=y/(1-y)$, one also finds
\begin{equation}
\label{phi(z)}
\varphi(z)=z^{-1}-(1-\rho)\sum_{r=0}^{\infty}\C{\rho\,r}{r}\frac{(-1)^{r}z^{r}}{r+1}\;.
\end{equation}
The first term takes into account the pole at $y=1$ of $y/(1-y)$, so that the contour $\Gamma$ encloses both poles of the integrand $y=1$ and $y=g^{-1}(z)$.\\\indent
At half filling, the summation over $r$ in (\ref{phi(z)}) and (\ref{psi(z)}) can be done explicitly. One finds
\begin{eqnarray}
\label{phi(z) rho=1/2}
&& \varphi(z)=-\frac{1}{2}+\frac{\sqrt{4+z^{2}}}{2z}\\
\label{psi(z) rho=1/2}
&& \psi(z)=-2\,\arcsinh(z/2)=-2\log\Big(\frac{z}{2}+\sqrt{1+\frac{z^{2}}{4}}\Big)\;.
\end{eqnarray}
The expressions (\ref{psi(z)}) and (\ref{phi(z)}) rely on the assumption that the solution $b$ of (\ref{b[k]}) is such that $\Re\,b>-\ell$ with $\ell$ defined in (\ref{l[rho]}), otherwise the expansion of $g^{-1}$ for small argument would not be convergent. The condition is satisfied if $\gamma$ is a large enough real positive number, in which case $b\simeq\gamma$. Besides, when $\gamma=0$, numerical checks on small system seem to indicate that (\ref{b[k]}) always has a solution inside the radius of convergence of the series in $\rme^{-b}$ for all the eigenstates, except the one with eigenvalue $0$. For the latter, it seems that (\ref{b[k]}) does not have a solution $b\in\mathbb{C}$, although formally $b=-\infty$ is a solution of (\ref{b[k]}) for which (\ref{e[k]}) gives $e=0$. We emphasize that for \textit{all} other solutions of (\ref{b[k]}), even the ones for which the eigenvalue is close to $0$ such as the gap, equation (\ref{b[k]}) seems to have a solution.
\begin{figure}
  \begin{tabular}{ccc}
    \begin{tabular}{c}\includegraphics[width=70mm]{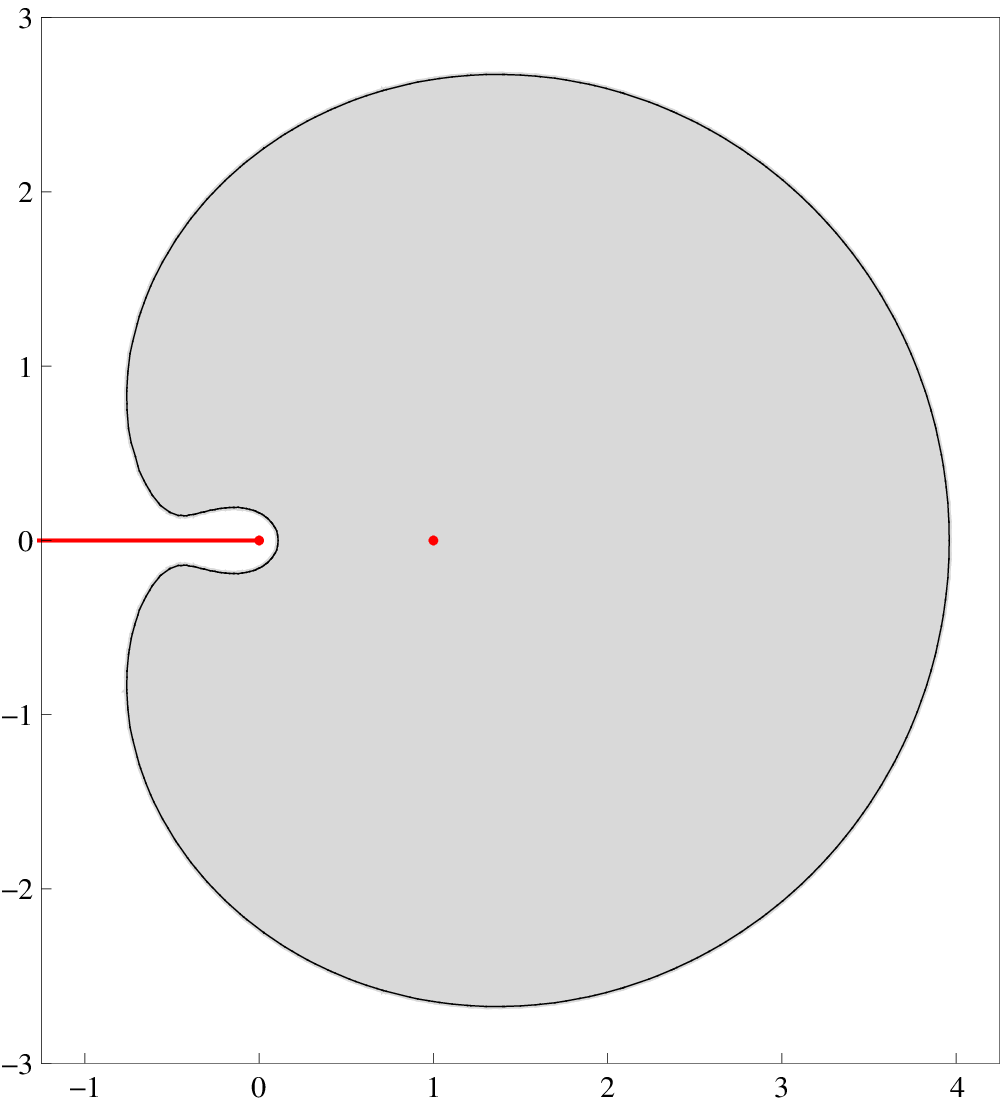}\end{tabular}
    & &
    \begin{tabular}{c}\includegraphics[width=70mm]{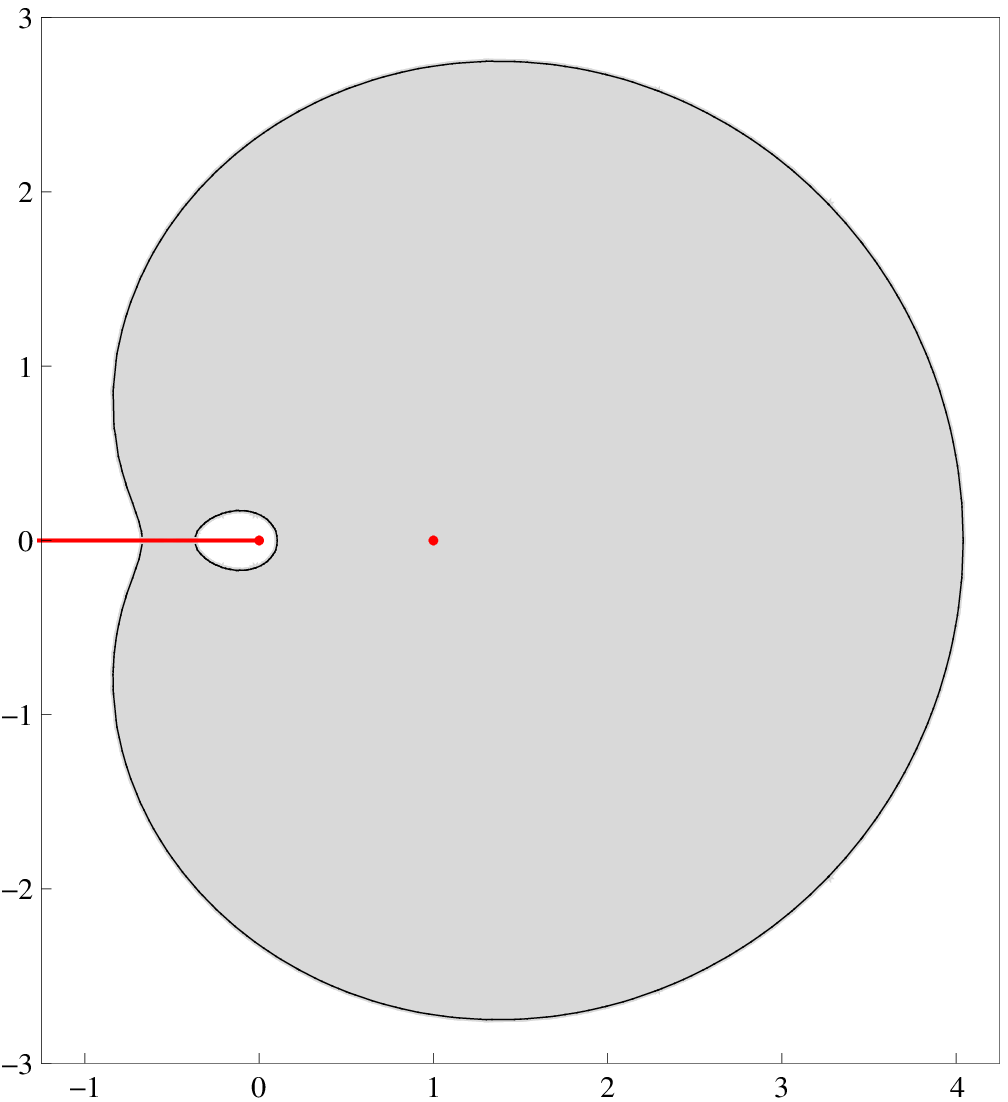}\end{tabular}
  \end{tabular}
  \caption{Curves of the points $y$ such that $|g(y)|=|z|$ for $\rho=1/3$ with $z=0.99\,\rme^{\ell}$ (left) and $z=1.01\,\rme^{\ell}$ (right). The function $g$ is defined in eq.~(\ref{g}) and $\ell$ is given by (\ref{l[rho]}). The gray area corresponds to $|g(y)|<|z|$ and the white area to $|g(y)|>|z|$. The red half line is the cut $\mathbb{R^{-}}$ of the function $g$.}
  \label{fig curves g}
\end{figure}
\end{subsection}

\begin{subsection}{Thermodynamic limit}
\label{subsection eigenvalues thermodynamic limit}
For large $L$, both (\ref{e[k]}) and (\ref{b[k]}) become independent of the detailed structure of the $k_{j}$'s, and only retain information about the density profile of the $k_{j}$'s. For each eigenstate, we introduce a function $\eta$ such that $L\,\eta(u)\rmd u$ is the number of $k_{j}$ in the interval $[L\,u,L(u+\rmd u)]$. The function $\eta$ then obeys the normalization
\begin{equation}
\label{rho[eta]}
\rho=\int_{0}^{1}\rmd u\,\eta(u)\;,
\end{equation}
while the eigenvalue (\ref{e[k]}) becomes
\begin{equation}
\label{e[eta]}
e=\int_{0}^{1}\rmd u\,\eta(u)\varphi\big(\rme^{2\rmi\pi u-b}\big)\;,
\end{equation}
and the equation for the parameter $b$ (\ref{b[k]}) rewrites
\begin{equation}
\label{b[eta]}
b=\gamma+\int_{0}^{1}\rmd u\,\eta(u)\psi\big(\rme^{2\rmi\pi u-b}\big)\;.
\end{equation}
In sections \ref{section density} and \ref{section trace}, we will need to count the number of eigenstates corresponding to a given function $\eta$ in order to sum over the eigenvalues. Since the number of ways to place $L\,\eta(u)\rmd u$ $k_{j}$'s in any interval $L\,\rmd u$ is $\C{L\,\rmd u}{L\,\eta(u)\rmd u}$, Stirling's formula implies that the total number of eigenstates corresponding to $\eta$ is $\Omega[\eta]\sim\rme^{Ls}$, where we defined an "entropy per site"
\begin{equation}
\label{s[eta]}
s=-\int_{0}^{1}\rmd u\,[\eta(u)\log\eta(u)+(1-\eta(u))\log(1-\eta(u))]\;.
\end{equation}
\end{subsection}
\end{section}

\begin{section}{Edge of the spectrum}
\label{section envelope}
In the thermodynamic limit, the rescaled eigenvalues $e=E/L$ fill a bounded domain in the complex plane, see fig. \ref{fig spectrum}. We study in this section the boundary of this domain, called in the following edge of the spectrum.\\\indent
We observe numerically that the eigenvalues located at the edge of the spectrum correspond to eigenstates for which the $k_{j}$'s are consecutive numbers. There are $L$ such possibilities, that we index with an integer $m$ between $0$ and $L-1$. We write $k_{j}=m+j-(N+1)/2$. The eigenvalue with largest real part corresponds to $m=0$ for $\gamma>0$. For $\gamma=0$, however, we remind that there is no solution to (\ref{b[k]}) when $m=0$. We will see that it corresponds to a singular point for the curve at the edge of the spectrum.\\\indent
In the limit $L\to\infty$, the corresponding density profile $\eta(u)$ of the $k_{j}$'s is the function with period $1$ such that
\begin{equation}
\eta(u)=\Big|
  \begin{array}{lll}
    1 && \text{for}\;\mu-\frac{\rho}{2}<u<\mu+\frac{\rho}{2}\\
    0 && \text{otherwise}
  \end{array}\;,
\end{equation}
where we defined $\mu=m/L$.
With this choice of $\eta$, writing explicitly the dependency in $\mu$ of the eigenvalue and of the parameter $b$, one finds
\begin{equation}
e(\mu)=\int_{-\rho/2}^{\rho/2}\rmd u\,\varphi\big(\rme^{2\rmi\pi(u+\mu)-b(\mu)}\big)\;
\end{equation}
and
\begin{equation}
b(\mu)=\int_{-\rho/2}^{\rho/2}\rmd u\,\psi\big(\rme^{2\rmi\pi(u+\mu)-b(\mu)}\big)\;.
\end{equation}
At half-filling, a nice parametric representation of the curve $e(\mu)$ can be written by replacing $b(\mu)$ by a new variable $d(\mu)$ defined by
\begin{equation}
\label{d[b]}
d(\mu)=-\rmi\arccos\Big(\frac{\rme^{2\rmi\pi\mu-b(\mu)}}{2}\Big)\;.
\end{equation}
Taking the derivative with respect to $\mu$ of the equation for $b(\mu)$ and calculating explicitly the integrals, one obtains
\begin{eqnarray}
\label{e[d]}
&& e(\mu)=-\frac{\tanh d(\mu)-d(\mu)}{2\rmi\pi}\\
\label{d'[d]}
&& d'(\mu)=\frac{\pi^{2}}{d(\mu)\tanh d(\mu)}\;.
\end{eqnarray}
The initial condition for the differential equation (\ref{d'[d]}) depends on the value of $\gamma$. For $\gamma=0$, since $e(0)=0$ (largest eigenvalue of a Markov matrix), then one must have $d(0)=0$, which corresponds to $b(0)=-\log2$. This is a singular point for the differential equation (\ref{d'[d]}), which is related to the fact that $b=-\log2$ corresponds to the border of the region where the expansion of $\varphi(z)$ and $\psi(z)$ for small $z$ in section \ref{section eigenvalues} is convergent.\\\indent
Expanding the differential equation at second order near $\mu=0$ leads to $3$ solutions. Inserting them in the equation for $e(\mu)$, one finds that only the solution
\begin{equation}
\label{d(mu) 0}
d(\mu)=\rme^{-2\rmi\pi/3}(3\pi^{2}\mu)^{1/3}+\frac{\pi^{2}\mu}{5}+\O{\mu^{5/3}}\;
\end{equation}
gives $\Re\,e(\mu)<0$ for small $\mu$ positive or negative. One has
\begin{equation}
\label{e(mu) 0}
e(\mu)=-\frac{\rmi\pi\mu}{2}+\frac{\rmi\rme^{2\rmi\pi/3}3^{2/3}\pi^{7/3}\mu^{5/3}}{10}+\O{\mu^{7/3}}\;.
\end{equation}
Studying the stability of the solutions of the differential equation near $d=0$, one observes that taking as initial condition $d(0)=\epsilon\,\rme^{\rmi\,\theta}$ with $0<\epsilon\ll1$ and $-\pi<\theta<-\pi/3$ leads to the correct solution, see fig.~\ref{fig stability d(0)}.\\\indent
We observe from (\ref{e(mu) 0}) that the curve $e(\mu)$ is singular near the origin with a power $5/3$ as announced in (\ref{Ree[Ime]}). In fig.~\ref{fig spectrum}, the curve $e(\mu)$ obtained from (\ref{e[d]}) and (\ref{d'[d]}) is plotted along with the full spectrum of TASEP for $L=18$, $N=9$. The agreement is already very good with the large $L$ limit.
\begin{figure}
  \begin{center}
    \includegraphics[width=70mm]{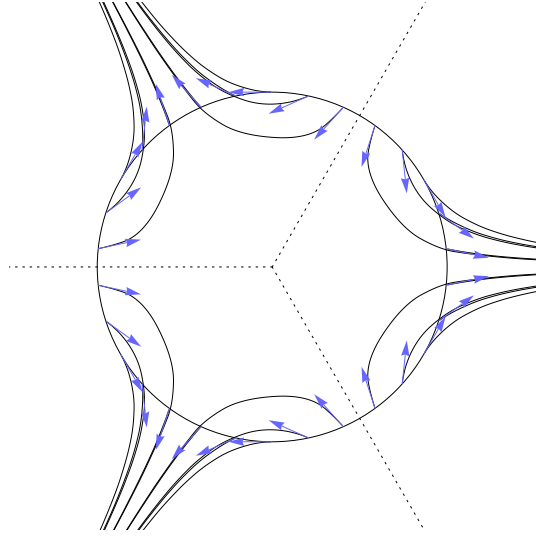}
  \end{center}
  \caption{Solutions of the differential equation (\ref{d'[d]}) for small $\mu$ with initial condition $d(0)=10^{-10}\rme^{\rmi\,\theta}$, for $\theta=\pi/30,3\pi/30,\ldots,59\pi/30$. The initial values $d(0)$ are represented on the circle. The arrows indicate the direction of the increment given by the derivative at $\mu=0$, $d'(\mu)\simeq\pi^{2}/d(0)^{2}$. The curves starting on the circle represent the solution of the differential equation (\ref{d'[d]}) solved numerically, which is essentially undistinguishable from the solution of $d'(\mu)=\pi^{2}/d(\mu)^{2}$ for the values of $d$ shown. The dotted lines represent the $3$ solutions of (\ref{d'[d]}) unstable under a perturbation near $d=0$.}
  \label{fig stability d(0)}
\end{figure}

\begin{subsection}{Dilogarithm}
An alternative expression to (\ref{d'[d]}) can be written by solving explicitly the differential equation for $d$ in terms of a dilogarithm function, as
\begin{equation}
\label{Li2[d]}
\fl\hspace{5mm} \frac{d(\mu)^{2}}{2}-\frac{\pi^{2}}{24}+d(\mu)\log\big(1+\rme^{-2d(\mu)}\big)-\frac{1}{2}\Li_{2}\big(-\rme^{-2d(\mu)}\big)=\pi^{2}\mu-\pi^{2}\Big(1+\Big\lfloor\mu-\frac{1}{5}\Big\rfloor\Big)\;.
\end{equation}
For $\mu$ in a neighbourhood of $0$, the rhs of the previous equation is equal to $\pi^{2}\mu$. The second term in the rhs is however needed since $\rme^{-2d(\mu)}$ crosses the cut $[1,\infty)$ of the dilogarithm at $\mu=1/5$, see fig.~\ref{fig exp(-2d)}. Indeed, inserting $d(1/5)=-\log((1+\sqrt{5})/2)-\rmi\pi/2$ and using
\begin{equation}
\Li_{2}\Big(\frac{3+\sqrt{5}}{2}\Big)=-\frac{11\pi^{2}}{15}-\log^{2}\Big(-\frac{1+\sqrt{5}}{2}\Big)\;,
\end{equation}
one observes that the equation (\ref{Li2[d]}) is verified, with $\rme^{-2d(1/5)}=(3+\sqrt{5})/2>1$. For $\mu=-1/5$, the solution is also explicit: using
\begin{equation}
\Li_{2}\Big(\frac{3-\sqrt{5}}{2}\Big)=\frac{\pi^{2}}{15}-\log^{2}\Big(\frac{1+\sqrt{5}}{2}\Big)\;,
\end{equation}
one finds $d(-1/5)=\log((1+\sqrt{5})/2)-\rmi\pi/2$. Here however, $\rme^{-2d(-1/5)}=(3-\sqrt{5})/2<1$ does not cross the cut of the dilogarithm.
\begin{figure}
  \begin{center}
    \begin{tabular}{ccc}
      \begin{tabular}{c}\includegraphics[width=50mm]{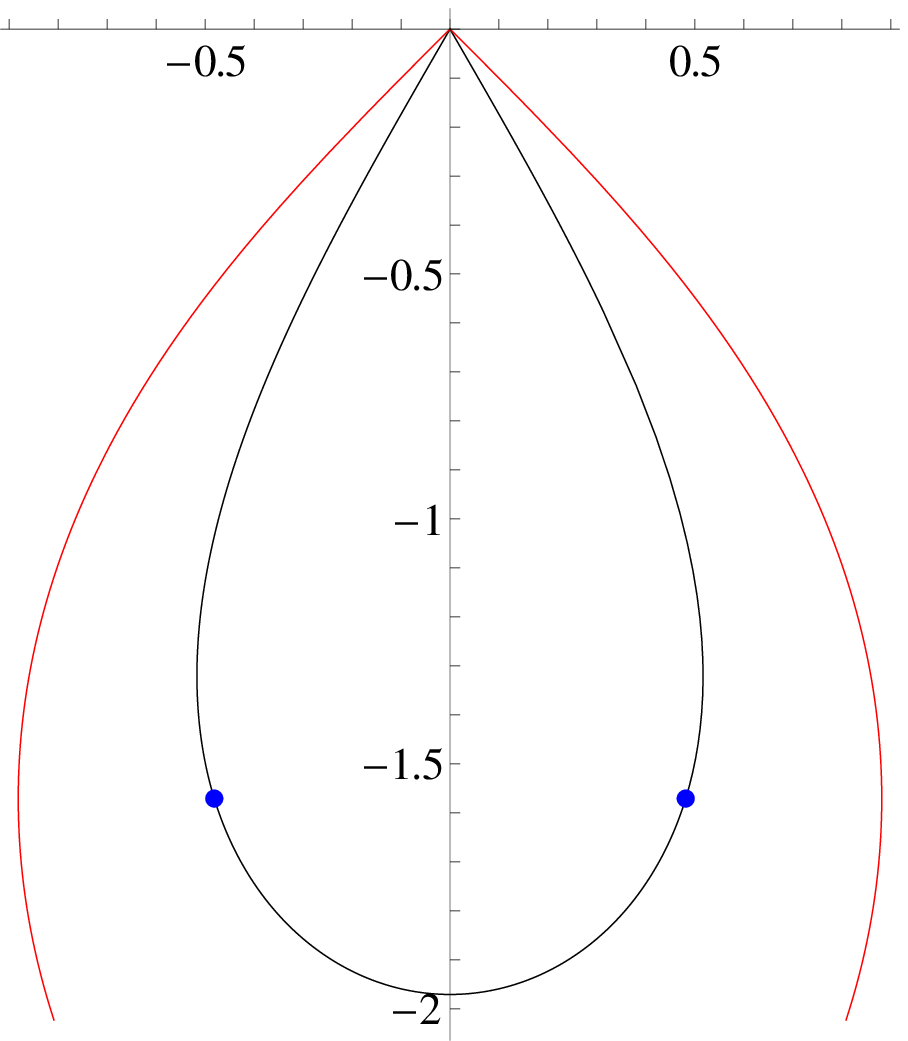}\end{tabular}
      & \qquad &
      \begin{tabular}{c}\includegraphics[width=70mm]{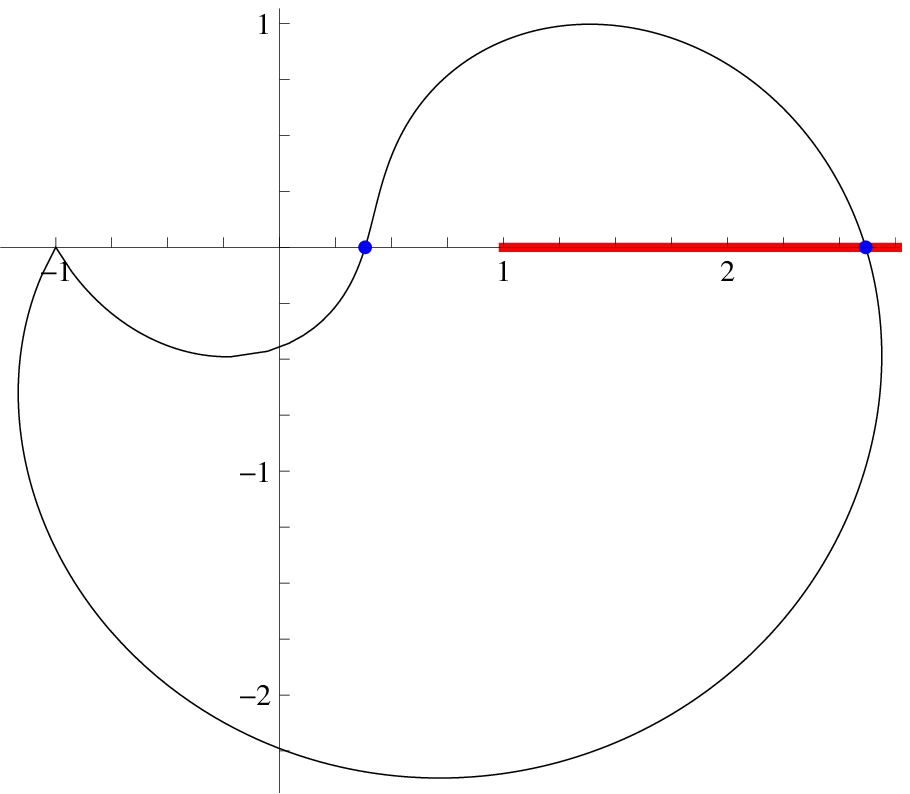}\end{tabular}
    \end{tabular}
  \end{center}
  \caption{Curve in the complex plane of $d(\mu)$ (left) and $-\rme^{-2d(\mu)}$ (right), from a numerical resolution of the differential equation (\ref{d'[d]}). In both graphs, the two dots are the explicit values $-\rme^{-2d(\pm1/5)}=(3\pm\sqrt{5})/2$. In the graph on the left, the outer, red curve encloses the domain for which $\Re\,b>-\log2$, for which the expansions (\ref{psi(z)}) and (\ref{phi(z)}) hold. In the graph on the right, the thick, red line correspond to the cut of the dilogarithm $[1,\infty)$.}
  \label{fig exp(-2d)}
\end{figure}
\begin{figure}
  \begin{center}
    \begin{tabular}{c}\includegraphics[width=100mm]{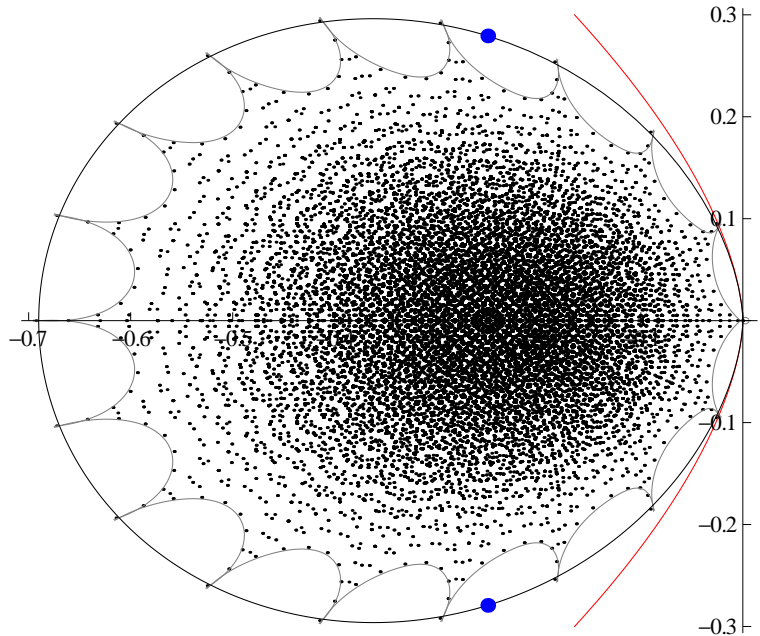}\end{tabular}
  \end{center}
  \caption{Spectrum of TASEP with $N=9$ particles on $L=18$ sites. The small black dots are the eigenvalues $E$ of the Markov matrix divided by $L$. The black curve is the edge of the spectrum in the thermodynamic limit at half-filling. The red (gray in printed version) curve corresponds to the asymptotic expression (\ref{Ree[Ime]}) of the edge of the spectrum near the origin. The two big blue dots correspond to the explicit values $E/L=-\frac{1}{4}\pm\frac{1}{2\rmi\pi}\big(\sqrt{5}-\log\frac{1+\sqrt{5}}{2}\big)$. The first order correction to the edge from eq.~(\ref{deltae[d]}) is plotted in gray.}
  \label{fig spectrum}
\end{figure}
\end{subsection}

\begin{subsection}{Edge of the spectrum, scale \texorpdfstring{$L^{0}$}{L^0} (half-filling)}
We observe in fig.~\ref{fig spectrum} that the edge of the spectrum shows $L$ small peaks. These peaks are a $1/L$ correction to the leading behaviour (\ref{e[d]}), (\ref{d'[d]}). They are a consequence of the constraint that all the integers $k_{j}$ are different. This phenomenon does not happen for non-interacting particles, see fig~\ref{fig s(e) free U}.\\\indent
In order to study this correction, one must go back to the exact expressions (\ref{e[k]}) and (\ref{b[k]}). Numerically, one observes that the $k_{j}$ that contribute to the edge are such that all the $k_{j}$'s are consecutive except at most one of them. We will write $k_{j}=\mu L+j-(N+1)/2$ for $j=1,\ldots,N-1$ and $k_{N}=(\mu+\nu)L$. On the scale studied here, $\mu$ can only take the values $1/L$, $2/L$, \ldots, $1$. The parameter $\nu$ verifies $0\leq\nu\leq1-\rho$. We focus again on the half-filled case $\rho=1/2$.\\\indent
At leading order in $L$, one recovers (\ref{e[d]}), (\ref{d'[d]}) using again the change of variable (\ref{d[b]}). Writing $e=e(\mu+1/(2L))+\delta e(\mu,\nu)$, one finds at the end of the calculation
\begin{equation}
\label{deltae[d]}
\hspace{-20mm}
\delta e(\mu,\nu)=\frac{\tanh d(\mu)}{2d(\mu)L}\Bigg(\frac{d(\mu)\sqrt{1-\rme^{4\rmi\pi\nu}\cosh^{2}d(\mu)}}{\rmi\,\rme^{2\rmi\pi\nu}\sinh d(\mu)}-\arcsinh\Big[\rmi\,\rme^{2\rmi\pi\nu}\cosh d(\mu)\Big]\Bigg)\;,
\end{equation}
where $d(\mu)$ is the solution of (\ref{d'[d]}).\\\indent
From (\ref{Li2[d]}) and the relation $\Li_{2}(-\rme^{z})+\Li_{2}(-\rme^{-z})=-\pi^{2}/6-z^{2}/2$, the function $d(\mu)$ verifies the symmetry relation $d(-\mu)=-\overline{d(\mu)}$, where $\overline{\,\cdot\,}$ denotes complex conjugation. This implies $\delta e(-\mu,1/2-\nu)=\overline{\delta e(\mu,\nu)}$. The latter symmetry is a consequence of the term $+1/(2L)$ in the definition of $\delta e$. The first-order correction (\ref{deltae[d]}) is plotted in fig.~\ref{fig spectrum} along with the exact spectrum for $L=18$, $N=9$.
\end{subsection}
\end{section}

\begin{section}{Density of eigenstates}
\label{section density}
The total number of eigenstates for TASEP is $|\Omega|\sim\rme^{\ell L}$, with $\ell$ defined in terms of the density of particles $\rho$ in (\ref{l[rho]}). In the bulk of the spectrum, the number of eigenstates with a rescaled eigenvalue $E/L$ close to a given $e$ is expected to be of the form $\rme^{Ls(e)}$. The function $s(e)$ is studied in this section.

\begin{subsection}{Optimal function \texorpdfstring{$\eta$}{eta}}
The density of eigenstates near the rescaled eigenvalue $e$ can be formally defined by the functional integral
\begin{equation}
\label{D[path integral]}
D(e)=\int\mathcal{D}\eta\,\openone_{\{e[\eta]=e\}}\openone_{\{\int_{0}^{1}\rmd u\,\eta(u)=\rho\}}\rme^{L s[\eta]}\;.
\end{equation}
We write explicitly the dependency in $\eta$ of $e$, $b$ and $s$ in this section. We want to maximize for $\eta$
\begin{equation}
\label{s + Lagrange multipliers}
s[\eta]+\lambda\Big(\int_{0}^{1}\rmd u\,\eta(u)-\rho\Big)+\Re(2\omega(e[\eta]-e))\;.
\end{equation}
It gives an optimal function $\eta^{*}$, which depends on the two Lagrange multipliers $\lambda\in\mathbb{R}$ and $\omega\in\mathbb{C}$. Those must then be set such that the constraints $\int_{0}^{1}\rmd u\,\eta(u)=\rho$ and $e[\eta^{*}]=e$ are satisfied, and we can finally write (\ref{D[s]}) with $s(e)=s[\eta^{*}]$.\\\indent
For given values of the Lagrange multipliers, writing the variation of $s[\eta]$, $b[\eta]$ and $e[\eta]$ for a small variation $\delta\eta$ of $\eta$ and using (\ref{rho[eta]}), (\ref{e[eta]}) and (\ref{b[eta]}), we find
\begin{equation}
\label{deltas[deltaeta]}
\delta s=-\int_{0}^{1}\rmd u\,\delta\eta(u)\log\frac{\eta(u)}{1-\eta(u)}\;,
\end{equation}
\begin{equation}
\label{deltab[deltaeta]}
\delta b=\frac{\int_{0}^{1}\rmd u\,\delta\eta(u)\psi\big(\rme^{2\rmi\pi u-b[\eta]}\big)}{1+\frac{1}{2\rmi\pi}\int_{0}^{1}\rmd u\,\eta(u)\partial_{u}\psi\big(\rme^{2\rmi\pi u-b[\eta]}\big)}\;,
\end{equation}
and
\begin{equation}
\label{deltae[deltaeta]}
\delta e=\int_{0}^{1}\rmd u\,\delta\eta(u)\Big[\varphi\big(\rme^{2\rmi\pi u-b[\eta]}\big)+a[\eta]\psi\big(\rme^{2\rmi\pi u-b[\eta]}\big)\Big]\;.
\end{equation}
We have defined
\begin{equation}
\label{a[eta]}
a[\eta]=-\frac{\frac{1}{2\rmi\pi}\int_{0}^{1}\rmd u\,\eta(u)\partial_{u}\varphi\big(\rme^{2\rmi\pi u-b[\eta]}\big)}{1+\frac{1}{2\rmi\pi}\int_{0}^{1}\rmd u\,\eta(u)\partial_{u}\psi\big(\rme^{2\rmi\pi u-b[\eta]}\big)}\;.
\end{equation}
It implies that the optimal function $\eta^{*}$ verifies
\begin{equation}
-\log\frac{\eta^{*}(u)}{1-\eta^{*}(u)}+\lambda+\Re[2\,\omega\,\varphi\big(\rme^{2\rmi\pi u-b[\eta^{*}]}\big)+2\,a[\eta^{*}]\,\omega\,\psi\big(\rme^{2\rmi\pi u-b[\eta^{*}]}\big)]\;.
\end{equation}
The optimal function is then equal to
\begin{equation}
\label{eta* density}
\eta^{*}(u)=\Big(1+\rme^{-\lambda-2\,\Re[\omega\,\varphi\big(\rme^{2\rmi\pi u-b[\eta^{*}]}\big)+a[\eta^{*}]\,\omega\,\psi\big(\rme^{2\rmi\pi u-b[\eta^{*}]}\big)]}\Big)^{-1}\;.
\end{equation}
This is a real function that satisfies $0<\eta^{*}(u)<1$ for all $u$. The expression (\ref{eta* density}) is reminiscent of a Fermi-Dirac distribution. On the other hand, the corresponding expression for undistinguishable non-interacting particles, (\ref{eta* free U}), resembles a Bose-Einstein distributions: allowing some momenta to be equal gives a term $-1$ in the denominator of (\ref{eta* free U}), while forbidding equal momenta gives the term $+1$ in the denominator of (\ref{eta* density}).
\end{subsection}

\begin{subsection}{Contour integrals}
For given values of the Lagrange multipliers $\lambda$ and $\omega$, the expression (\ref{eta* density}) is completely explicit except for the two unknown complex quantities $a[\eta^{*}]$ and $b[\eta^{*}]$, that must be determined self-consistently from (\ref{a[eta]}) and (\ref{b[eta]}). It is possible to simplify the problem a little by noticing that we can replace $a[\eta^{*}]$ and $b[\eta^{*}]$ by two real quantities $\alpha$ and $\beta$. Indeed, the definitions (\ref{a[eta]}) and (\ref{eta* density}) imply
\begin{eqnarray}
&& \Im(a[\eta^{*}]\omega)=\Im\Big[-\frac{\omega}{2\rmi\pi}\int_{0}^{1}\rmd u\,\eta^{*}(u)\partial_{u}\Big(\varphi\big(\rme^{2\rmi\pi u-b}\big)+a\psi\big(\rme^{2\rmi\pi u-b}\big)\Big)\Big]\nonumber\\
&&\hspace{19mm} =\frac{1}{4\pi}\int_{0}^{1}\rmd u\,\frac{(\eta^{*})'(u)}{1-\eta^{*}(u)}=0\;.
\end{eqnarray}
We define $\alpha=a[\eta^{*}]\omega\in\mathbb{R}$. The imaginary part of $b[\eta^{*}]$ can be eliminated by a shift of $u$ and a redefinition of $\eta$: we introduce $\sigma$ such that
\begin{equation}
\sigma(\rme^{2\rmi\pi u})=\eta^{*}\Big(u+\frac{\Im(b)}{2\pi}\Big)\;.
\end{equation}
Defining $\beta=\Re(b[\eta^{*}])$, one has
\begin{equation}
\label{sigma(z) density}
\sigma(z)=\Big(1+\rme^{-\lambda\,-2\,\Re[\omega\varphi(\rme^{-\beta}z)]\,-2\alpha\,\Re[\psi(\rme^{-\beta}z)]}\Big)^{-1}\;.
\end{equation}
It is not possible to eliminate the quantity $\beta$ by changing the contour of integration, since \textit{e.g.} $\Re[\psi(\rme^{-\beta}z)]$ is not an analytic function of $z$. It can also be seen by writing $2\,\Re[\psi(\rme^{-\beta}z)]=\psi(\rme^{-\beta}z)+\psi(\rme^{-\beta}z^{-1})$ for $|z|=1$.\\\indent
The quantities $s=s[\eta^{*}]$, $\beta$, $\alpha$ can be rewritten in terms of $\sigma$ as
\begin{equation}
\label{s[sigma]}
s=-\frac{1}{2\rmi\pi}\oint\frac{\rmd z}{z}\,\sigma(z)\log\sigma(z)+(1-\sigma(z))\log(1-\sigma(z))\;,
\end{equation}
\begin{equation}
\label{beta[sigma] density}
\beta=\Re\,\gamma+\Re\Big[\frac{1}{2\rmi\pi}\oint\frac{\rmd z}{z}\,\sigma(z)\psi\big(\rme^{-\beta}z\big)\Big]\;,
\end{equation}
and
\begin{equation}
\label{alpha[sigma] density}
\alpha=-\frac{1}{2\rmi\pi}\oint\rmd z\,\sigma(z)\partial_{z}\Big(\omega\varphi\big(\rme^{-\beta}z\big)+\alpha\psi\big(\rme^{-\beta}z\big)\Big)\;.
\end{equation}
All the contour integrals are over the circle of radius $1$ and center $0$ in the complex plane. Similarly, the constraints for $\rho$ and $e[\eta^{*}]$ give
\begin{equation}
\label{rho[sigma]}
\rho=\frac{1}{2\rmi\pi}\oint\frac{\rmd z}{z}\,\sigma(z)\;,
\end{equation}
and
\begin{equation}
\label{e[sigma] density}
e=\frac{1}{2\rmi\pi}\oint\frac{\rmd z}{z}\,\sigma(z)\varphi\big(\rme^{-\beta}z\big)\;.
\end{equation}
For given $\rho$ and $e$, one has to solve (\ref{rho[sigma]}) and (\ref{e[sigma] density}) in order to obtain $\lambda$ and $\omega$ in terms of them. Like for (\ref{alpha[sigma] density}) and (\ref{beta[sigma] density}), it does not seem that these equations can be solved analytically in general. At half-filling, however, it is possible to show that
\begin{equation}
\label{lambda(omega)}
\lambda=\Re\,\omega\;,
\end{equation}
leaving only the equations for $\alpha$, $\beta$ and $\omega$ to be solved numerically. Indeed, for $\rho=1/2$, one has the identities $\varphi(-z)=-1-\varphi(z)$ and $\psi(-z)=-\psi(z)$. Setting $\lambda=\Re\,\omega$ in (\ref{sigma(z) density}) then implies $\sigma(z)+\sigma(-z)=1$, from which (\ref{rho[sigma]}) follows at half-filling.\\\indent
We note that if $\omega\in\mathbb{R}$ then $e\in\mathbb{R}$. This is a consequence of (\ref{e[sigma] density}), $\sigma(-z)=\sigma(z)$ and $\Im[\varphi(-z)]=-\Im[\varphi(z)]$. By the definition (\ref{s + Lagrange multipliers}) of the Lagrange multiplier $\omega$, it implies that $s$ is also the density of eigenvalues with a given real part when $\omega$ is real.\\\indent
The maximum of $s$ is located at $e=-\rho(1-\rho)$, $s=\ell$. It corresponds to $\omega=0$, $\sigma(z)=\rho$, $\lambda=\log[\rho/(1-\rho)]$, $\alpha=0$, $\beta=\Re\,\gamma$. Unlike free particles, the spectrum is not symmetric with respect to the maximum of $s$.\\\indent
In fig.~\ref{fig e(omega)}, $e$ is plotted for various values of $\omega$ at half-filling, along with the optimal function $\sigma(\rme^{2\rmi\pi u})$. In fig.~\ref{fig s(e)}, $s$ is plotted as a function of $e\in\mathbb{R}$ and compared with the density of real part of eigenvalues obtained from numerical diagonalization of the Markov matrix $M$ for $N=9$, $L=18$. The agreement is not very good for eigenvalues close to the edges. This is caused by the "arches" at distance $\sim1/L$ of the edge of the spectrum, which still contribute much for $N=9$, $L=18$, see fig.~\ref{fig spectrum}.
\begin{figure}
  \begin{center}
    \begin{tabular}{ccc}
      \begin{tabular}{c}\includegraphics[width=70mm]{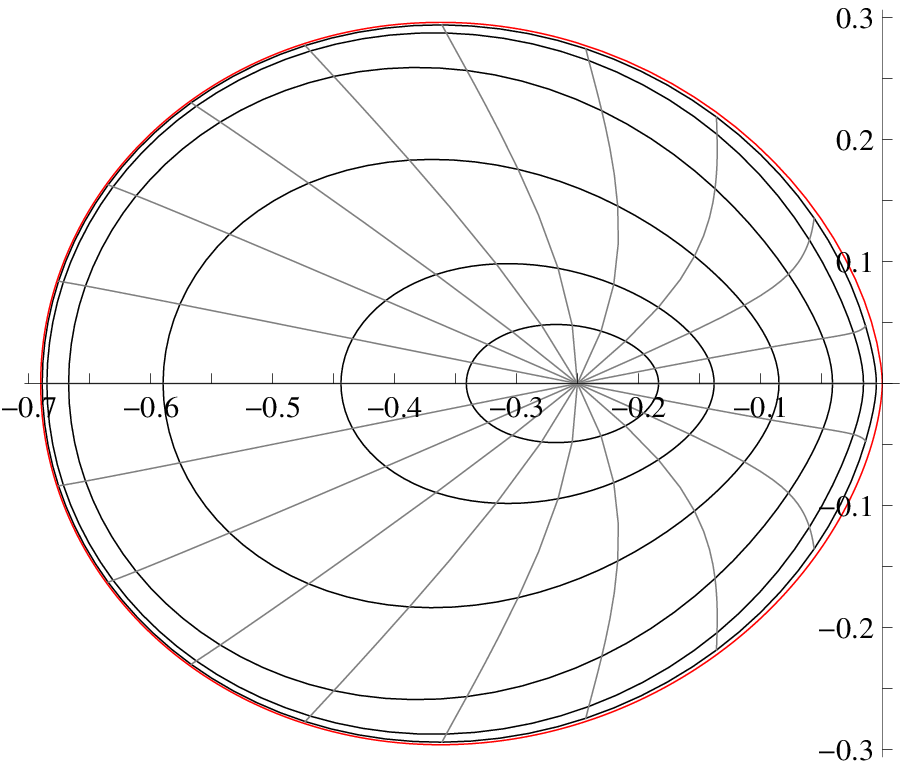}\end{tabular}
      &&
      \begin{tabular}{c}\includegraphics[width=70mm]{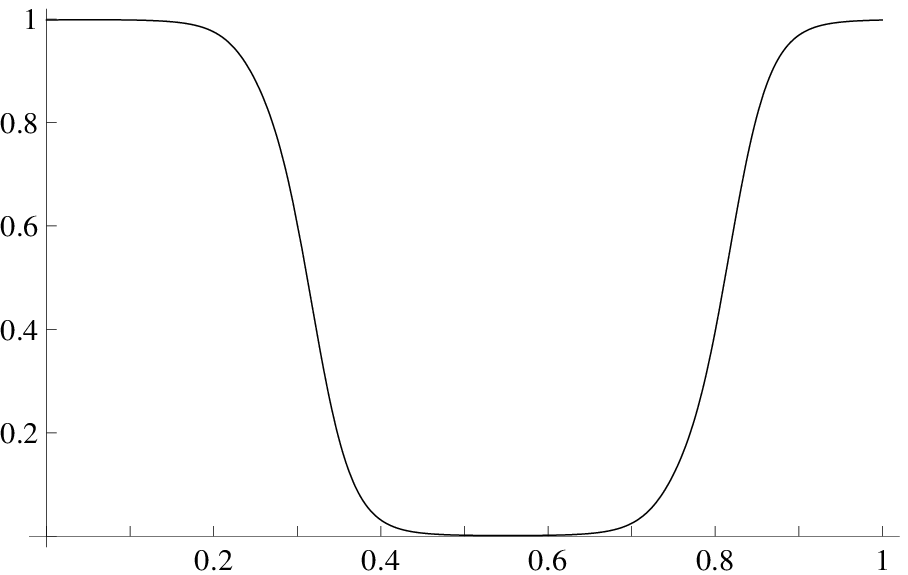}\end{tabular}
    \end{tabular}
  \end{center}
  \caption{On the left, graph of $e(\omega)$ for fixed values of $|\omega|$ (black), and fixed values of $\arg\omega$ (gray), obtained from (\ref{e[sigma] density}) after solving numerically the system (\ref{beta[sigma] density}), (\ref{alpha[sigma] density}) at half-filling (left). The different curves correspond to $|\omega|=0.25,0.5,1,2,4,8$ (from the center to the edge) and to $\arg\omega=0,\pi/10,2\pi/10,\ldots,19\pi/20$. The outer, red curve is the edge of the spectrum, computed numerically from (\ref{e[d]}), which is recovered from (\ref{e[sigma] density}) in the limit $|\omega|\to\infty$. On the right, optimal function $\sigma(\rme^{2\rmi\pi u})$ plotted as a function of $u$ for $\omega=8\,\rme^{\rmi\pi/10}$.}
  \label{fig e(omega)}
\end{figure}
\begin{figure}
  \begin{center}
    \begin{tabular}{c}\includegraphics[width=70mm]{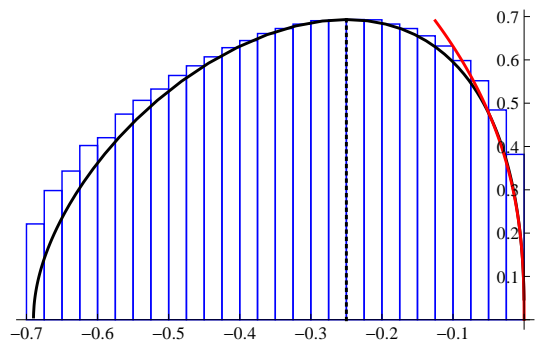}\end{tabular}
  \end{center}
  \caption{Number of eigenvalues with real part $L\,e$ of the Markov matrix $M$ of TASEP at half-filling, plotted as a function of $e$. The thick black curve corresponds to the expressions (\ref{s[sigma]}), (\ref{e[sigma] density}) parametrized by $\omega$ ranging from $-50$ to $60$, where the quantities $\alpha$ and $\beta$ are solved numerically using equations (\ref{alpha[sigma] density}) and (\ref{beta[sigma] density}) for each value of the parametrization $\omega$. The thick red (gray in printed version) curve is the asymptotics (\ref{s(e)}). The histograms correspond to the density of real part of eigenvalues obtained from numerical diagonalization of $M$ for the finite system with $N=9$ particles on $L=18$ sites. Because of the logarithmic corrections in $L$ of $L^{-1}\log|\Omega|\simeq\log2$, where $|\Omega|=\C{L}{L/2}$ is the total number of microstates, we shift the height of the histograms so that their maximum is $\log2$.}
  \label{fig s(e)}
\end{figure}
\end{subsection}

\begin{subsection}{Density of eigenvalues close to the origin (\texorpdfstring{$\rho=1/2$}{rho=1/2})}
\label{subsection density e=0}
We consider the limit $e\to0$ of $s(e)$ at half-filling for the Markov matrix (deformation $\gamma=0$). It corresponds to $|\omega|\to\infty$ with $\arg\omega\to0$.\\\indent
In the limit $|\omega|\to\infty$, the optimal function $\sigma(\rme^{2\rmi\pi u})$ approaches $1$ if $u_{1}<u<u_{2}$ and $0$ if $u_{2}<u<u_{1}+1$. The relation $\sigma(-z)=1-\sigma(z)$ at half-filling and the normalization condition (\ref{rho[sigma]}) imply that $u_{2}=u_{1}+1/2$. Furthermore, if we consider a scaling such that $\arg\omega\to0$ when $|\omega|\to\infty$, one has $u_{1}=-1/4$ and $u_{2}=1/4$, which is the same as what we had in section \ref{section envelope} for the edge of the spectrum near the eigenvalue $0$.\\\indent
In \ref{appendix density}, we compute explicitly the large $\omega$ limit of the integrals (\ref{s[sigma]}), (\ref{beta[sigma] density}), (\ref{alpha[sigma] density}) and (\ref{e[sigma] density}) with $\lambda$ given by (\ref{lambda(omega)}). We find two different regimes, depending on the respective scaling between the real and imaginary part of $e$.\\\indent
The first regime, $e\to0$ with $|\Im\,e|^{5/3}/\Re\,e\to0$, corresponds to the central part of the spectrum, far from the edge. We find $\beta\simeq-\log2+\deltabeta|\omega|^{-2/3}$, where $\deltabeta$ is the solution of
\begin{equation}
\label{deltabeta c=0}
\frac{1}{6\pi}=\int_{0}^{\infty}\rmd x\,\frac{\Im[(1+2\rmi\pi x)^{1/2}]}{1+\rme^{\frac{(2\deltabeta)^{3/2}}{3}\,\Im[(1+2\rmi\pi x)^{3/2}]}}\;.
\end{equation}
The real part of the eigenvalue $e$ and the "entropy" $s$ are equal to
\begin{equation}
\label{Re(e) c=0}
\Re\,e=\frac{4\sqrt{2}\,\deltabeta^{5/2}}{3|\omega|^{5/3}}\Big[\frac{1}{10\pi}-\int_{0}^{\infty}\rmd x\,\frac{\Im[(1+2\rmi\pi x)^{3/2}]}{1+\rme^{\frac{(2\deltabeta)^{3/2}}{3}\,\Im[(1+2\rmi\pi x)^{3/2}]}}\Big]\;
\end{equation}
and
\begin{equation}
\label{s c=0}
s=-5\,|\omega|\,\Re\,e\;.
\end{equation}
Solving (\ref{deltabeta c=0}) numerically, one finds $\deltabeta\simeq0.706532$, which implies $\Re\,e\simeq-0.147533|\omega|^{-5/3}$, $s\simeq0.737667|\omega|^{-2/3}$ and (\ref{s(e)}).\\\indent
The second regime, $e\to0$ with $\Re\,e$ and $\Im\,e$ related by (\ref{Ree[Ime]}), corresponds to the edge of the spectrum. In this regime, one finds
\begin{equation}
s(e)\simeq\frac{2^{4/5}\pi^{2/5}}{3^{3/10}5^{1/10}}(-\Re\,e)^{2/5}\sqrt{1+\frac{2^{2/3}3^{7/6}\pi^{2/3}}{10}\,\frac{|\Im\,e|^{5/3}}{\Re\,e}}\;.
\end{equation}
The crossover between the two regimes corresponds to $|\Im\,e|^{5/3}/(-\Re\,e)$ converging to a constant different from the $2^{2/3}3^{7/6}\pi^{2/3}/10$ characteristic of the edge. Explicit expressions are given in \ref{appendix density}.
\end{subsection}
\end{section}

\begin{section}{Trace of the time evolution operator}
\label{section trace}
In this section, we study the quantity $f(t)$ defined in eq.~(\ref{f[M]}). This is another application of the formulas (\ref{e[eta]}), (\ref{b[eta]}) derived in section \ref{section eigenvalues} for the eigenvalues.

\begin{subsection}{Optimal function \texorpdfstring{$\eta$}{eta}}
As in section \ref{section density} for the density of eigenvalues, one can write the summation over all eigenvalues as
\begin{equation}
\label{tr[path integral]}
\tr\rme^{tM}=\sum_{\{k_{1},\ldots,k_{N}\}}\rme^{tE(k_{1},\ldots,k_{N})}\simeq\int\mathcal{D}\eta\,\openone_{\{\int_{0}^{1}\rmd u\,\eta(u)=\rho\}}\rme^{L(s[\eta]+te[\eta])}\;.
\end{equation}
If the functional integral is dominated by the contribution of an optimal function $\eta^{*}$, one finds from (\ref{f[M]}) $f(t)=s[\eta^{*}]+te[\eta^{*}]$. The function $\eta^{*}$ generally depends on $t$. The normalization (\ref{rho[eta]}) of $\eta$ is enforced by the Lagrange multiplier
\begin{equation}
\lambda\Big(-\rho+\int_{0}^{1}\rmd u\,\eta(u)\Big)\;.
\end{equation}
The change in $s$, $b$ and $e$ from a variation $\delta\eta$ of $\eta$ is still given by (\ref{deltas[deltaeta]}), (\ref{deltab[deltaeta]}) and (\ref{deltae[deltaeta]}). The optimal function then verifies
\begin{equation}
-\log\frac{\eta^{*}(u)}{1-\eta^{*}(u)}+\lambda+t\Big(\varphi\big(\rme^{2\rmi\pi u-b[\eta^{*}]}\big)+a[\eta^{*}]\psi\big(\rme^{2\rmi\pi u-b[\eta^{*}]}\big)\Big)\;,
\end{equation}
with $a[\eta]$ still given by (\ref{a[eta]}). We obtain
\begin{equation}
\label{eta* tr}
\eta^{*}(u)=\Big(1+\rme^{-\lambda-t\big(\varphi\big(\rme^{2\rmi\pi u-b[\eta^{*}]}\big)+a[\eta^{*}]\psi\big(\rme^{2\rmi\pi u-b[\eta^{*}]}\big)\big)}\Big)^{-1}\;.
\end{equation}
\textit{Remark}: unlike section \ref{section density}, the optimal function $\eta^{*}(u)$ is not a real function. This means that the saddle point of the functional integral (\ref{tr[path integral]}) lies in the complex plane. The function $\eta^{*}$ of (\ref{eta* tr}) can be recovered from the function $\eta^{*}$ of (\ref{eta* density}) by the choice $\omega=t$ and $\overline{\omega}=0$, where $\overline{\,\cdot\,}$ denotes complex conjugation, and $\omega$ and $\overline{\omega}$ must be thought of as independent variables.
\end{subsection}

\begin{subsection}{Contour integrals}
The optimal function $\eta^{*}(u)$ is an analytic function of $z=\rme^{2\rmi\pi u-b[\eta^{*}]}$. One defines
\begin{equation}
\label{sigma(z) trace 0}
\sigma(z)=\eta^{*}(u)=\Big(1+\rme^{-\lambda-t(\varphi(z)+a[\eta^{*}]\psi(z))}\Big)^{-1}\;.
\end{equation}
Under the assumption that the contours of integration can be freely deformed from $|z|=\rme^{-\Re\,b[\eta^{*}]}$ to something independent of $b[\eta^{*}]$, we recover the equations (\ref{s[sigma]}) and (\ref{rho[sigma]}) for $s\equiv s[\eta^{*}]$ and $\rho$, but with $\sigma$ now given by (\ref{sigma(z) trace 0}) instead of (\ref{sigma(z) density}). The equations for the quantities $e\equiv e[\eta^{*}]$ and $a=a[\eta^{*}]$ become
\begin{equation}
\label{e[sigma] tr}
e=\frac{1}{2\rmi\pi}\oint\frac{\rmd z}{z}\,\sigma(z)\varphi\big(z\big)\;
\end{equation}
and
\begin{equation}
\label{a[sigma] tr}
a=-\oint\frac{\rmd z}{2\rmi\pi}\,\sigma(z)\,\partial_{z}\big(\varphi(z)+a\,\psi(z)\big)\;.
\end{equation}
Eq.~(\ref{sigma(z) trace 0}) implies
\begin{equation}
\varphi(z)+a\psi(z)=-\frac{\lambda}{t}-\frac{1}{t}\log\frac{1-\sigma(z)}{\sigma(z)}\;.
\end{equation}
Combining this with (\ref{a[sigma] tr}) gives
\begin{equation}
a=\oint\frac{\rmd z}{2\rmi\pi t}\,\partial_{z}\log(1-\sigma(z))=\frac{w}{t}\;,
\end{equation}
where $w\in\mathbb{Z}$ is the winding number of $1-\sigma(z)$ around the origin. We assume in the following that $w=0$, hence $a=0$ and
\begin{equation}
\label{sigma(z) trace}
\sigma(z)=\Big(1+\rme^{-\lambda-t\varphi(z)}\Big)^{-1}\;.
\end{equation}
The property $a=0$ is compatible, at least for small times, with numerical solutions of (\ref{rho[sigma]}) with $a=0$, see fig.~\ref{fig sigma}.\\\indent
Unlike section \ref{section density}, the expression (\ref{sigma(z) trace}) for $\sigma(z)$ allows to calculate explicitly the Lagrange multiplier $\lambda$, the eigenvalue $e$, the "entropy" $s$ and the function $f(t)=s+t\,e$. Details are given in \ref{appendix trace}. In the end, we recover (\ref{f(t)}).
\end{subsection}

\begin{subsection}{Singularities of \texorpdfstring{$\sigma(z)$}{sigma(z)}}
\begin{figure}
  \begin{center}
    \begin{tabular}{ccc}
      \begin{tabular}{c}\includegraphics[width=60mm]{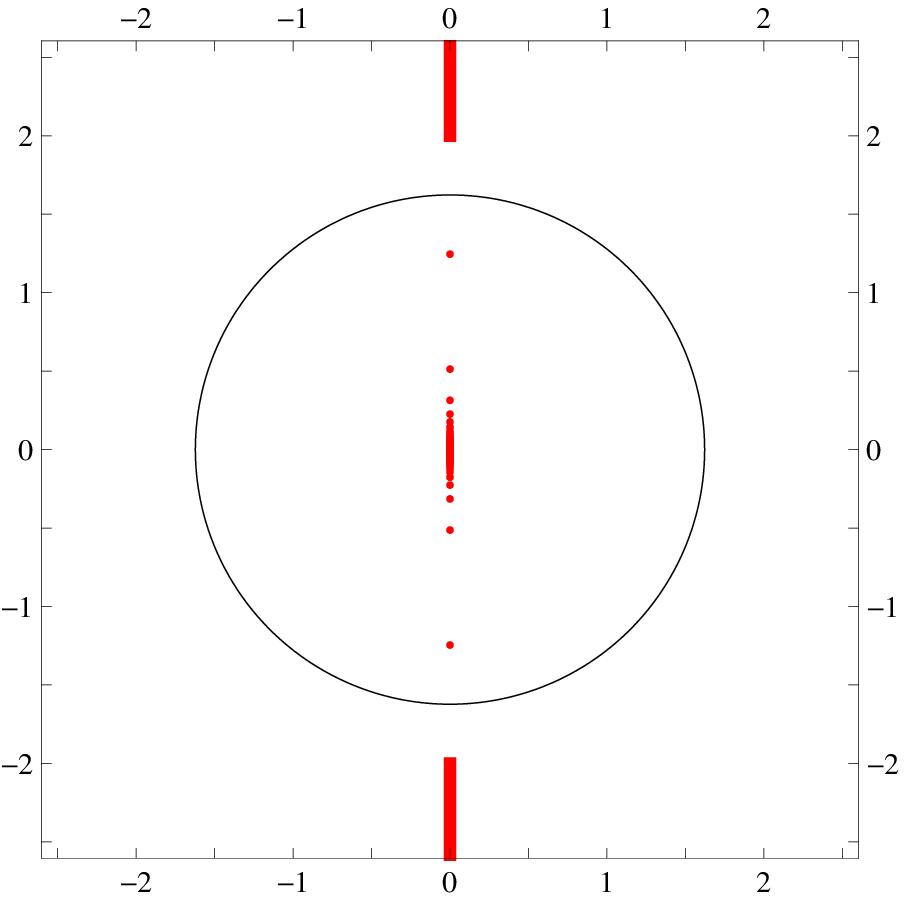}\end{tabular}
      &&
      \begin{tabular}{c}\includegraphics[width=50mm]{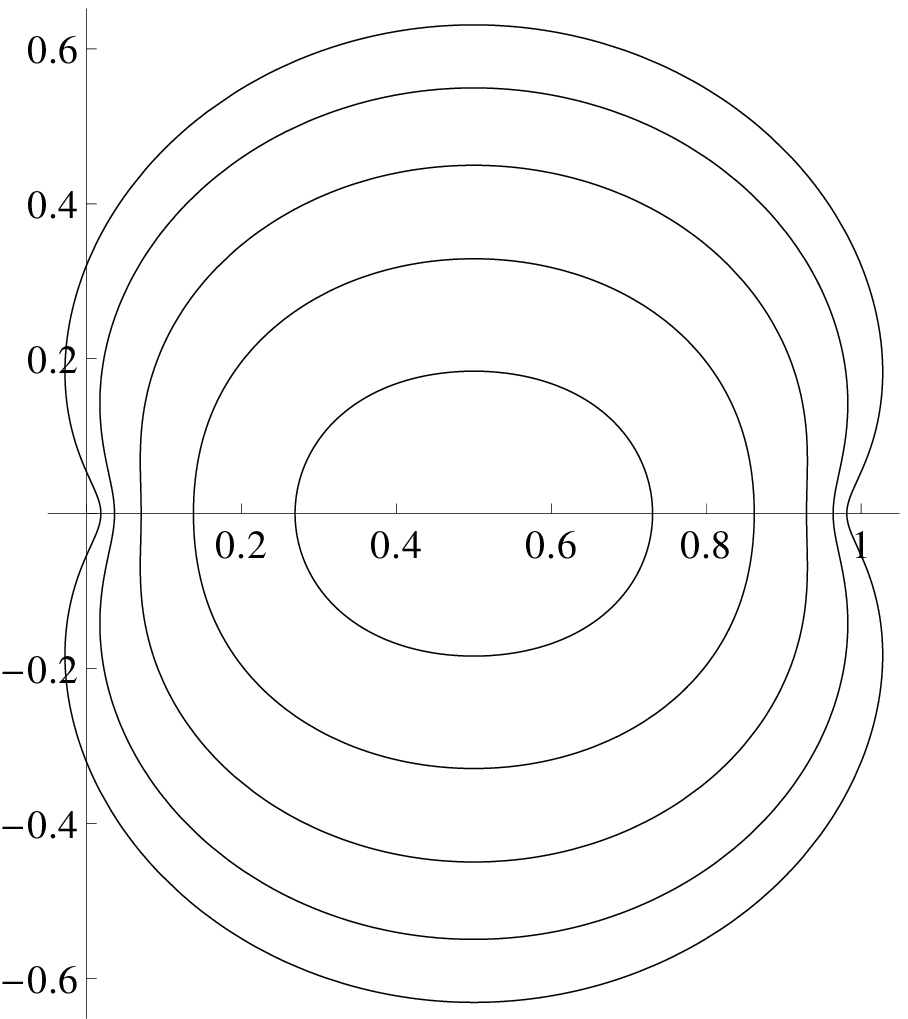}\end{tabular}
    \end{tabular}
  \end{center}
  \caption{On the left, the singularities of the function $\sigma$ for $t=5$ (poles and cuts) are drawn in red (gray in printed version). The circle of radius $r_{t}$ (\ref{r[t]}) represents a possible contour of integration. On the right, the image by $\sigma$ of this contour is drawn for $t$ between $1$ (inner curve) to $5$ (outer curve).}
  \label{fig sigma}
\end{figure}
In the previous subsection, we used a deformation of the contour of integration in order to eliminate completely the parameter $b$. We implicitly assumed that this was possible without crossing singularities of $\sigma(z)$. We come back to this issue here, with the expression (\ref{sigma(z) trace}) for $\sigma(z)$ resulting from the assumption $a=0$.\\\indent
We focus on the half-filled case. From (\ref{phi(z) rho=1/2}) and (\ref{lambda(t,rho)}), the singularities of the function $\sigma$ consist in an essential singularity at $z=0$, two cuts starting at $\pm2\rmi$, and an infinity of poles located at $\pm z_{k}$, $k\in\mathbb{N}$, with
\begin{equation}
z_{k}=\frac{2\rmi t}{\sqrt{t^{2}+4(2k+1)^{2}\pi^{2}}}\;.
\end{equation}
For $t=0$, we observe that all the poles are located at the origin. For times $t<t(\gamma)$, where $t(\gamma)$ is the first non-analyticity of $f$ discussed at the end of section \ref{subsection Q}, the contour of integration in (\ref{s[sigma]}) and (\ref{e[sigma] tr}) then has to enclose all the poles, but must not cross the cuts. The circle of center $0$ and radius 
\begin{equation}
\label{r[t]}
r_{t}=1+t/\sqrt{t^{2}+4\pi^{2}}\;
\end{equation}
is a possible contour.\\\indent
We observed in section \ref{subsection Q} that the function $f(t)$ is defined piecewise. The non-analyticity of $f(t)$ should be the sign of the presence of several saddle points competing in the functional integral (\ref{tr[path integral]}). This is similar to what happens in the direct calculations of $f(t)$ for free particles in \ref{appendix free particles}, with a simple integral instead of a functional integral. It is not completely clear, however, how several saddle points emerge from the functional integral (\ref{tr[path integral]}) for TASEP. It might be due to a change in the contour of integration at $t=t(\gamma)$, with a new contour that does not enclose all the poles of $\sigma$. There could also be a transition in eq.~(\ref{a[sigma] tr}) from the solution $a=0$ to another value due to a change in the winding number of $1-\sigma(z)$ around $0$.
\end{subsection}
\end{section}

\begin{section}{Conclusion}
Parametric expressions can be derived for all the eigenvalues of TASEP using Bethe ansatz. These expressions allow a study of large scale properties of the spectrum in the thermodynamic limit, in particular the curve marking the edge of the spectrum, the density of eigenvalues in the bulk of the spectrum and the generating function of the cumulants of the eigenvalues.\\\indent
A natural extension of the present work would be to analyse the structure of the eigenvalues closer to the origin. Of particular interest are eigenvalues with a real part scaling as $L^{-3/2}$, which control the relaxation to the stationary state. Another goal would be to obtain asymptotic expressions for the scalar product between an eigenstate characterized by a density $\eta$ of $k_{j}$'s as in sections \ref{section density} and \ref{section trace} and a microstate characterized by a density profile of particles. This would allow to study physical quantities more interesting than the trace of the time evolution operator.\\\indent
Another very interesting extension would be the case of the asymmetric simple exclusion process with partial asymmetry, where particles are allowed to hop in both directions, with an asymmetry parameter controlling the bias. It would be nice if it were possible to derive parametric expressions for the eigenvalues analogous to (\ref{e[k]}) and (\ref{b[k]}). A good starting point seems to be the quantum Wronskian formulation of the Bethe equations \cite{PS99.1,P10.1}, where a parameter analogous to the parameter $B$ we used here exists. It would allow to study how the spectrum changes at the transitions between equilibrium and non-equilibrium.\\\indent
Finally, it would be nice to understand how the approach used here to study the spectrum of TASEP relates to thermodynamic Bethe ansatz \cite{YY69.1}. The latter follows from the observation that, in the thermodynamic limit, Bethe roots accumulate on a curve in the complex plane. The density of Bethe roots along this curve can usually be shown to be the solution of a nonlinear integral equation. We would like to understand whether the absence of non-linear integral equations in our calculations is only due to the special decoupling (\ref{polynomial[y,B]=0}) of the Bethe equations for TASEP.

\begin{subsection}*{Acknowledgements}
I thank B. Derrida for several very helpful discussions. I also thank D. Mukamel for his warm welcome at the Weizmann Institute of Science, where early stages of this work were done.
\end{subsection}
\end{section}

\appendix
\begin{section}{Bijection \texorpdfstring{$g$}{g}}
\label{appendix g}
The function $g$ is defined in (\ref{g}) on the whole complex plane minus the negative real axis $(-\infty,0]$, if one chooses the usual cut of the logarithm. It is convenient to use polar coordinates, writing $y=r\,\rme^{\rmi\,\theta}$ with $r>0$ and $-\pi<\theta<\pi$. Then $g(y)$ is divergent in the limit of small and large values of $r$. One has
\begin{eqnarray}
&& g(y)\simeq r^{-\rho}\rme^{-\rmi\,\rho\,\theta}\quad\text{for}\;r\to0\\
&& g(y)\simeq-r^{1-\rho}\rme^{\rmi(1-\rho)\theta}\quad\text{for}\;r\to\infty\;.
\end{eqnarray}
This implies $\arg g(y)\in(-\rho\pi,\rho\pi)$ for small $r$ and $\arg g(y)\in(-\pi,-\rho\pi)\cup(\rho\pi,\pi]$ for large $r$. The image of a point on the cut of $g$ is
\begin{equation}
g(-r\pm\rmi\,0^{\pm})=\rme^{\pm\rmi\pi\rho}\,\frac{1+r}{r^{\rho}}\;.
\end{equation}
The derivative of $g(y)$ with respect to $\theta$ at this point verifies
\begin{equation}
\frac{\partial_{\theta}g}{g}(-r\pm\rmi\,0^{\pm})=\frac{\rmi(1-\rho)}{r+1}\Big[r-\frac{\rho}{1-\rho}\Big]\;.
\end{equation}
This implies that the curves $\{g(r\rme^{\rmi\,\theta}),\theta\in(-\pi,\pi)\}$ join the cuts in the image space orthogonally (which already follows from the local holomorphicity of $g$, as the image of the orthogonality of any circle of center $0$ with the negative real axis), from one side or the other depending on whether $r$ is smaller or larger than $\rho/(1-\rho)$.\\\indent
When $y$ spans $(-\infty,0]\pm0\,\rmi$, the image of the cut spans $\rme^{\rmi\pi\rho}[\rme^{\ell},\infty)\,\cup\,\rme^{-\rmi\pi\rho}[\rme^{\ell},\infty)$, with $\ell$ defined in equation (\ref{l[rho]}). The bijective nature of $g$ is clearly seen in fig.~\ref{fig grid g} and fig.~\ref{fig grid g-1} where the functions $g$ and its inverse $g^{-1}$ are represented.
\begin{figure}
  \begin{center}
    \begin{tabular}{ccc}
      \begin{tabular}{c}\includegraphics[width=70mm]{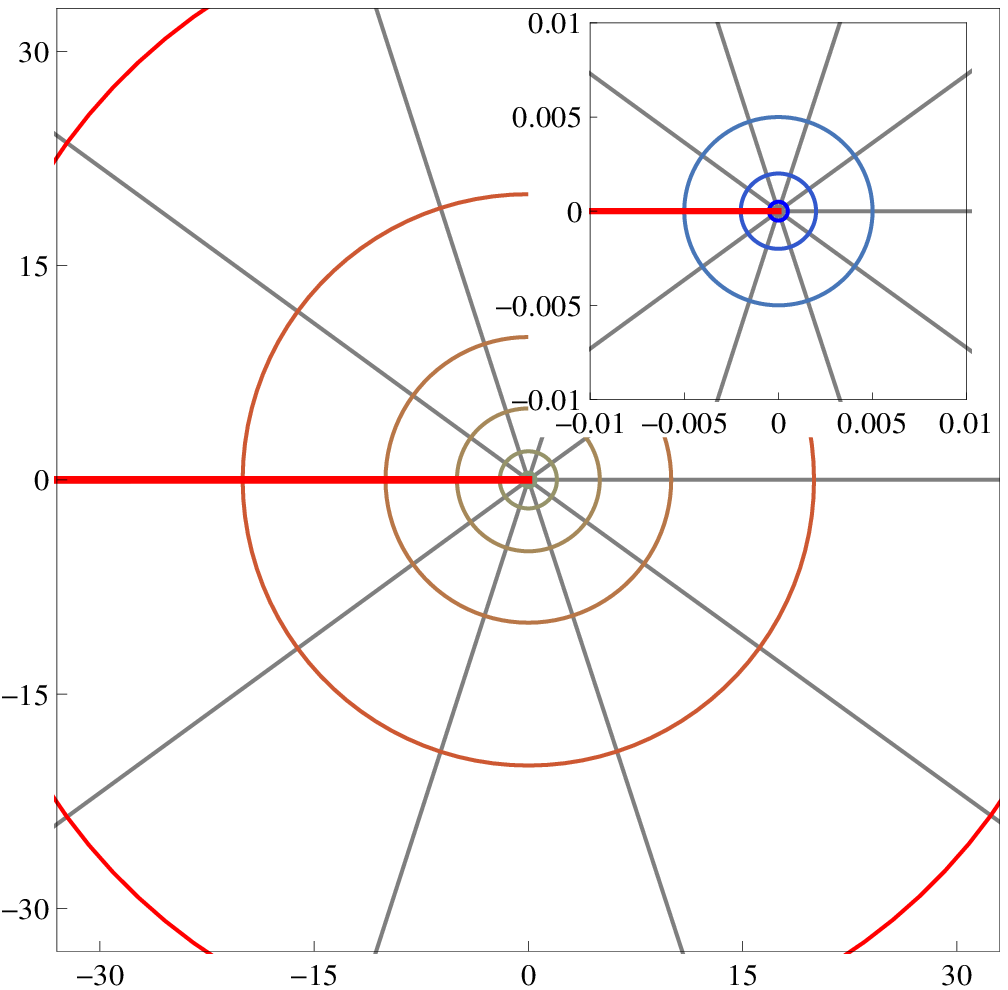}\end{tabular}
      &&
      \begin{tabular}{c}\includegraphics[width=70mm]{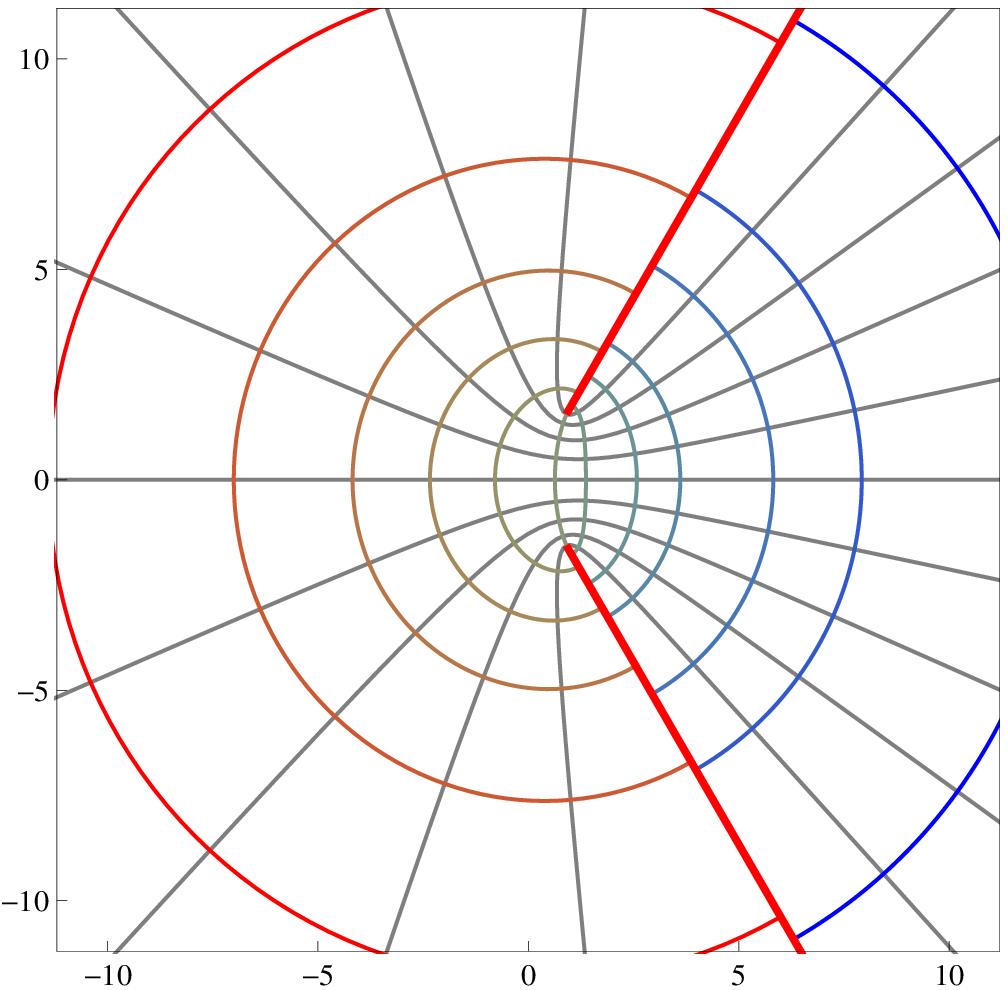}\end{tabular}
    \end{tabular}
  \end{center}
  \caption{Deformation of a grid by the function $g$ (\ref{g}), with $\rho=1/3$. On the left is a grid in polar coordinates, with angles $\theta$ regularly spaced of $2\pi/10$ and radii $0.0005$, $0.002$, $0.005$, $0.02$, $0.05$, $0.2$, $0.5$, $2$, $5$, $10$, $20$, $40$ coloured from blue for small radius to red for large radius. The thick red line corresponds to the cut of the function $g$. On the right, the image by the function $g$ of the previous grid is drawn, using the same colors for a curve in the initial grid and its image by $g$. The almost semi-circular curves on the left part of the complex plane are red, the ones on the right are blue. The two thick red lines correspond to the image of the cut by $g$.}
  \label{fig grid g}
\end{figure}
\begin{figure}
  \begin{center}
    \begin{tabular}{ccc}
      \begin{tabular}{c}\includegraphics[width=70mm]{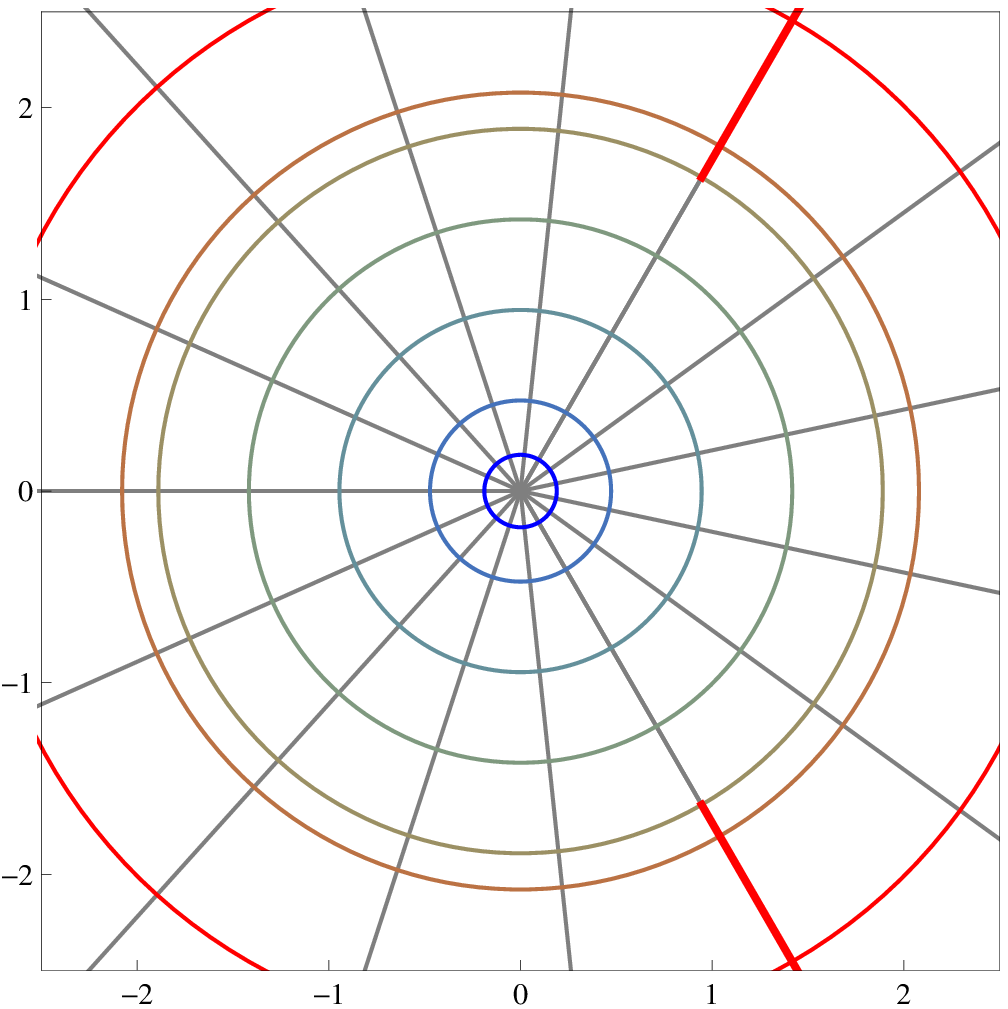}\end{tabular}
      &&
      \begin{tabular}{c}\includegraphics[width=70mm]{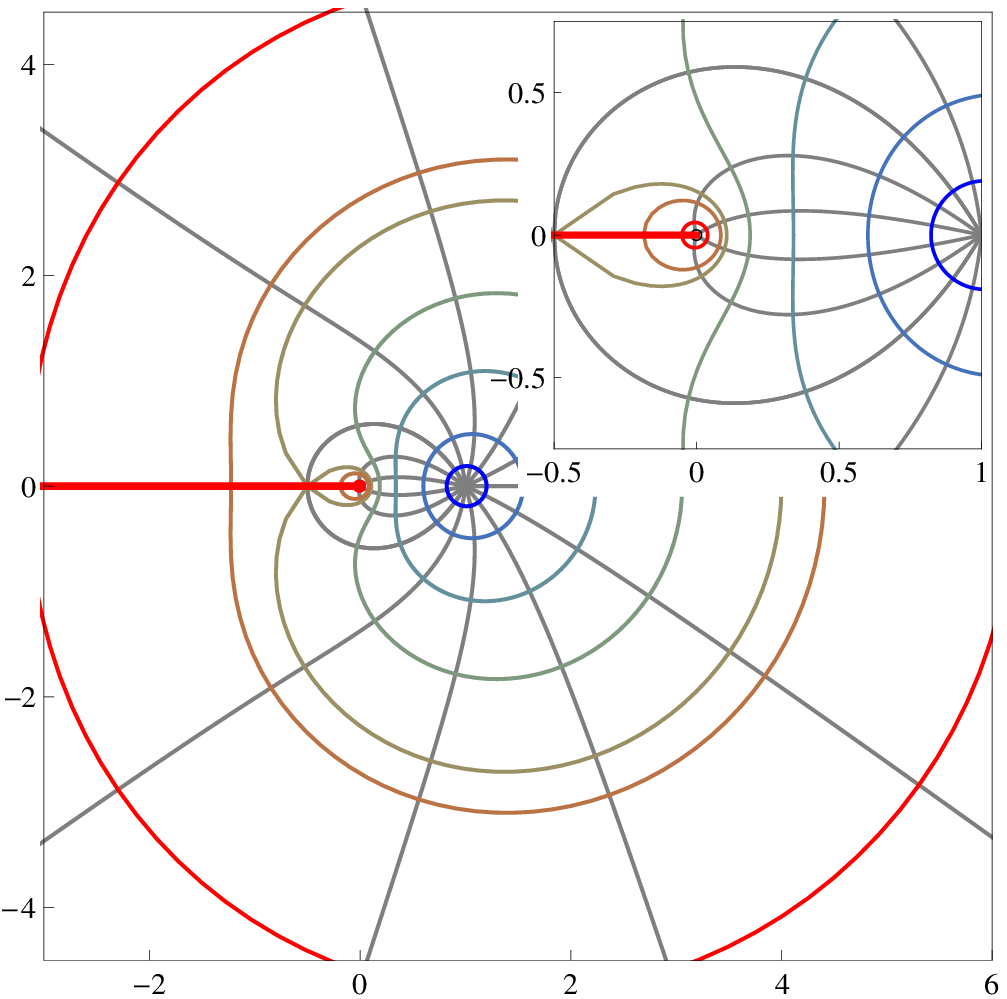}\end{tabular}
    \end{tabular}
  \end{center}
  \caption{Deformation of a grid by the inverse function $g^{-1}$ (\ref{g}), with $\rho=1/3$. On the left is a grid in polar coordinates, with angles $\theta$ regularly spaced of $2\pi/15$ and radii $0.1\,\rme^{\ell}$, $0.25\,\rme^{\ell}$, $0.5\,\rme^{\ell}$, $0.75\,\rme^{\ell}$, $\rme^{\ell}$, $1.1\,\rme^{\ell}$, $1.5\,\rme^{\ell}$ coloured from blue for small radius to red for large radius, with $\ell$ defined in (\ref{l[rho]}). The two thick red lines correspond to the cuts of the function $g^{-1}$. On the right is represented the image by the function $g^{-1}$ of the previous grid, using the same colors for a curve in the initial grid and its image by $g^{-1}$. The almost circular shapes of large radius and small radius around $0$ are red. The small almost circular shapes around $1$ are blue.}
  \label{fig grid g-1}
\end{figure}
\end{section}

\begin{section}{Density of eigenvalues close to the origin}
\label{appendix density}
In this appendix, we perform the calculations related to the limit $e\to0$ of the density of eigenvalues of the Markov matrix at half-filling. We compute the limit $|\omega|\to\infty$, $\arg\omega\to0$ of the integrals for the quantities $\alpha$ (\ref{alpha[sigma] density}), $\beta$ (\ref{beta[sigma] density}), $e$ (\ref{e[sigma] density}) and $s$ (\ref{s[sigma]}). The integrals must be decomposed as a sum of $4$ terms. Up to terms exponentially small in $\epsilon^{-1}\gg1$, one has
\begin{eqnarray}
&& \int_{0}^{1}\rmd u\,F[\sigma(\rme^{2\rmi\pi u}),\rme^{2\rmi\pi u}]\simeq\int_{-\frac{1}{4}+\epsilon}^{\frac{1}{4}-\epsilon}\rmd u\,F[1,\rme^{2\rmi\pi u}]+\int_{\frac{1}{4}+\epsilon}^{\frac{3}{4}-\epsilon}\rmd u\,F[0,\rme^{2\rmi\pi u}]\\
&&\hspace{25mm} +\int_{-\frac{1}{4}+\epsilon}^{-\frac{1}{4}-\epsilon}\rmd u\,F[\sigma(\rme^{2\rmi\pi u}),\rme^{2\rmi\pi u}]+\int_{\frac{1}{4}+\epsilon}^{\frac{1}{4}-\epsilon}\rmd u\,F[\sigma(\rme^{2\rmi\pi u}),\rme^{2\rmi\pi u}]\;.\nonumber
\end{eqnarray}
After a little rewriting, one obtains
\begin{eqnarray}
\label{intF}
&&\fl\hspace{5mm} \int_{0}^{1}\rmd u\,F[\sigma(\rme^{2\rmi\pi u}),\rme^{2\rmi\pi u}]\simeq\int_{-\frac{1}{4}}^{\frac{1}{4}}\rmd u\,F[1,\rme^{2\rmi\pi u}]+\int_{\frac{1}{4}}^{\frac{3}{4}}\rmd u\,F[0,\rme^{2\rmi\pi u}]\\
&&\fl\hspace{4mm} +\int_{-\epsilon}^{0}\rmd u\,\Big(F[\sigma(\rmi\,\rme^{2\rmi\pi u}),\rmi\,\rme^{2\rmi\pi u}]+F[\sigma(-\rmi\,\rme^{2\rmi\pi u}),-\rmi\,\rme^{2\rmi\pi u}]-F[1,\rmi\,\rme^{2\rmi\pi u}]-F[0,-\rmi\,\rme^{2\rmi\pi u}]\Big)\nonumber\\
&&\fl\hspace{4mm} +\int_{0}^{\epsilon}\rmd u\,\Big(F[\sigma(\rmi\,\rme^{2\rmi\pi u}),\rmi\,\rme^{2\rmi\pi u}]+F[\sigma(-\rmi\,\rme^{2\rmi\pi u}),-\rmi\,\rme^{2\rmi\pi u}]-F[0,\rmi\,\rme^{2\rmi\pi u}]-F[1,-\rmi\,\rme^{2\rmi\pi u}]\Big)\;.\nonumber
\end{eqnarray}
For the four quantities $\alpha$, $\beta$, $e$ and $s$, the terms with $F[0,\rme^{2\rmi\pi u}]$ vanishes, as well as the terms with $F[1,\rme^{2\rmi\pi u}]$ for $s$.\\\indent
In order to continue the calculations, several scalings need to be considered for $\Im\,\omega$ when $|\omega|\to\infty$. We write $\omega=r+\rmi\chi r^{c}$ with $r>0$, $\chi>0$, $c<1$, and will take the limit $r\to\infty$. The case $\chi<0$ then follows from the invariance of the spectrum by complex conjugation. A summary of the different scalings obtained is given in table \ref{table scalings}. The scalings $c=-2$, $-1$, $-1/2$, $0$, $1/6$, $1/5$, $1/4$, $1/3$, $1/2$, $2/3$, $3/4$, $4/5$ were checked by solving numerically (\ref{alpha[sigma] density}), (\ref{beta[sigma] density}) for $r=100,200,\ldots,1000$. We used the BST algorithm \cite{HS88.1} in order to improve the convergence to $r\to\infty$. For all the scalings studied, the relative errors in the numerical coefficients of $\alpha$, $\beta$, $\Re\,e$, $\Im\,e$ and $s$ obtained from the BST algorithm were smaller than $10^{-3}$ compared to the exact values.
\begin{table}
  \begin{center}
    $\displaystyle
    \begin{array}{cccccc}
      & \alpha-r/4 & \beta+\log2 & \Re\,e & \Im\,e & s\\
      &&&&&\\
      c<0 & -\frac{0.164473}{r^{1/3}} & \frac{0.706532}{r^{2/3}} & -\frac{0.147533}{r^{5/3}} & \frac{0.224251\,\chi}{r^{-c+4/3}} & \frac{0.737667}{r^{2/3}}\\
      & & \rotatebox{90}{$=$} & \rotatebox{90}{$=$} & & \rotatebox{90}{$=$}\\
      0<c<\frac{1}{3} & \frac{0.107018\,\chi^{2}}{r^{-2c+1/3}} & \frac{0.706532}{r^{2/3}} & -\frac{0.147533}{r^{5/3}} & \frac{0.224251\,\chi}{r^{-c+4/3}} & \frac{0.737667}{r^{2/3}}\\
      &&&&&\\
      \frac{1}{3}<c<1 & \frac{\sqrt{3}\chi r^{c}}{12} & \frac{\sqrt{3}\,\chi}{2r^{1-c}} & -\frac{2^{1/2}3^{3/4}\,\chi^{5/2}}{5\pi r^{5(1-c)/2}} & -\frac{2^{1/2}\,\chi^{3/2}}{3^{1/4}\pi r^{3(1-c)/2}} & \frac{\pi}{2^{1/2}3^{3/4}\chi^{1/2}r^{(1+c)/2}}
    \end{array}$
  \end{center}
  \caption{Various scalings of the parameters needed for the calculation of the density of eigenvalues close to the eigenvalue $0$, after writing $\omega=r+\rmi\chi r^{c}$. Exact expressions for the numerical constants are given in section \ref{subsection density e=0} and \ref{subsection scaling c=0}.}
  \label{table scalings}
\end{table}

\begin{subsection}{Scaling \texorpdfstring{$c=0$}{c=0}.}
\label{subsection scaling c=0}
Writing $\alpha=r/4+\deltaalpha/r^{1/3}$, $\beta=-\log2+\deltabeta/r^{2/3}$ and expanding the equations (\ref{beta[sigma] density}) and (\ref{alpha[sigma] density}) respectively up to order $r^{-1}$ and $r^{-2/3}$ with $\epsilon\sim r^{-2/3}$ in (\ref{intF}) give after straightforward, but rather tedious calculations (\ref{deltabeta c=0}) and
\begin{eqnarray}
\label{deltaalpha c=0}
&& \deltaalpha\Big[\frac{1}{2\pi}-\int_{0}^{\infty}\rmd x\,\frac{\Im[(1+2\rmi\pi x)^{-1/2}]}{1+\rme^{\frac{(2\deltabeta)^{3/2}}{3}\,\Im[(1+2\rmi\pi x)^{3/2}]}}\Big]\\
&& =\frac{\deltabeta^{2}}{3}\Big[\frac{1}{10\pi}-\int_{0}^{\infty}\rmd x\,\frac{\Im[(1+2\rmi\pi x)^{3/2}]}{1+\rme^{\frac{(2\deltabeta)^{3/2}}{3}\,\Im[(1+2\rmi\pi x)^{3/2}]}}\Big]\nonumber\\
&&\hspace{-7mm} +\frac{\sqrt{2\deltabeta}\,\chi^{2}}{4}\,\int_{0}^{\infty}\rmd x\,\frac{\Re[(1+2\rmi\pi x)^{1/2}]\,\Re[(1+2\rmi\pi x)^{-1/2}]\,\rme^{\frac{(2\deltabeta)^{3/2}}{3}\,\Im[(1+2\rmi\pi x)^{3/2}]}}{\Big(1+\rme^{\frac{(2\deltabeta)^{3/2}}{3}\,\Im[(1+2\rmi\pi x)^{3/2}]}\Big)^{2}}\;.\nonumber
\end{eqnarray}
We used the expansion
\begin{equation}
\int_{-1/4}^{1/4}\!\rmd u\,\psi(2\,\rme^{2\rmi\pi u-\epsilon})=-\log2+\epsilon-\frac{4\sqrt{2}}{3\pi}\,\epsilon^{3/2}+\frac{2\sqrt{2}}{15\pi}\,\epsilon^{5/2}+\O{\epsilon^{7/2}}\;.
\end{equation}
We observe that the equations (\ref{deltabeta c=0}) and (\ref{deltaalpha c=0}) for $\deltabeta$ and $\deltaalpha$ decouple in the scaling $c=0$, unlike the original equations (\ref{beta[sigma] density}) and (\ref{alpha[sigma] density}) for $\beta$ and $\alpha$.\\\indent
Expanding the equation (\ref{e[sigma] density}) for $e$, we find at leading order in $r$ (\ref{Re(e) c=0}) and
\begin{equation}
\label{Im(e) c=0}
\Im\,e=-\frac{4\deltabeta^{2}\chi}{r^{4/3}}\,\int_{0}^{\infty}\rmd x\,\frac{[\Re[(1+2\rmi\pi x)^{1/2}]]^{2}\,\rme^{\frac{(2\deltabeta)^{3/2}}{3}\,\Im[(1+2\rmi\pi x)^{3/2}]}}{\Big(1+\rme^{\frac{(2\deltabeta)^{3/2}}{3}\,\Im[(1+2\rmi\pi x)^{3/2}]}\Big)^{2}}\;,
\end{equation}
while the equation (\ref{s[sigma]}) for $s$ gives (\ref{s c=0}). Numerically, one has $\deltaalpha\simeq-0.164473+0.107018\,\chi^{2}$ and $\Im\,e\simeq-0.224251\,\chi\,r^{-4/3}$.
\end{subsection}

\begin{subsection}{Scaling \texorpdfstring{$c=1/3$}{c=1/3}.}
Writing $\alpha=r/4+\deltaalpha\,r^{1/3}$, $\beta=-\log2+\deltabeta\,r^{-2/3}$ and expanding the equations for $\beta$ (\ref{beta[sigma] density}) and $\alpha$ (\ref{alpha[sigma] density}) respectively up to order $r^{-1}$ and $r^{0}$ with $\epsilon\sim r^{-2/3}$ in (\ref{intF}) give after again long but straightforward calculations
\begin{equation}
\label{deltabeta c=1/3}
\frac{1}{3\pi}=\int_{0}^{\infty}\rmd x\,\Big[\frac{\Im[(1+2\rmi\pi x)^{1/2}]}{1+\rme^{\Phi_{+}(x)}}+\frac{\Im[(1+2\rmi\pi x)^{1/2}]}{1+\rme^{\Phi_{-}(x)}}\Big]\;
\end{equation}
and
\begin{eqnarray}
\label{deltaalpha c=1/3}
&& \deltaalpha\Big(\frac{1}{\pi}-\int_{0}^{\infty}\rmd x\,\Big[\frac{\Im[(1+2\rmi\pi x)^{-1/2}]}{1+\rme^{\Phi_{+}(x)}}+\frac{\Im[(1+2\rmi\pi x)^{-1/2}]}{1+\rme^{\Phi_{-}(x)}}\Big]\Big)\nonumber\\
&&\hspace{10mm} =-\frac{\chi}{4}\int_{0}^{\infty}\rmd x\,\Big[\frac{\Re[(1+2\rmi\pi x)^{-1/2}]}{1+\rme^{\Phi_{+}(x)}}-\frac{\Re[(1+2\rmi\pi x)^{-1/2}]}{1+\rme^{\Phi_{-}(x)}}\Big]\;,
\end{eqnarray}
with the definition
\begin{eqnarray}
&& \Phi_{\pm}(x)=\frac{(2\deltabeta)^{3/2}}{3}\,\Im[(1+2\rmi\pi x)^{3/2}]-4\sqrt{2\deltabeta}\,\deltaalpha\,\Im[(1+2\rmi\pi x)^{1/2}]\nonumber\\
&&\hspace{20mm} \pm\sqrt{2\deltabeta}\,\chi\,\Re[(1+2\rmi\pi x)^{1/2}]\;.
\end{eqnarray}
We observe that the equations for $\deltaalpha$ and $\deltabeta$ are now coupled, unlike in the scaling $c=0$. Similar calculations with the equations for $e$ (\ref{e[sigma] density}) and $s$ (\ref{s[sigma]}) give at leading order in $r$
\begin{equation}
\fl\hspace{15mm} \Re\,e=\frac{2\sqrt{2}\,\deltabeta^{5/2}}{3r^{5/3}}\Big(\frac{1}{5\pi}-\int_{0}^{\infty}\rmd x\,\Big[\frac{\Im[(1+2\rmi\pi x)^{3/2}]}{1+\rme^{\Phi_{+}(x)}}+\frac{\Im[(1+2\rmi\pi x)^{3/2}]}{1+\rme^{\Phi_{-}(x)}}\Big]\Big)\;,
\end{equation}
\begin{equation}
\fl\hspace{15mm} \Im\,e=\frac{\sqrt{2}\,\deltabeta^{3/2}}{r}\int_{0}^{\infty}\rmd x\,\Big[\frac{\Re[(\deltabeta+2\rmi\pi x)^{1/2}]}{1+\rme^{\Phi_{+}(x)}}-\frac{\Re[(1+2\rmi\pi x)^{1/2}]}{1+\rme^{\Phi_{-}(x)}}\Big]\;,
\end{equation}
and
\begin{equation}
\label{s c=1/3}
s=-5\,r\,\Re\,e+3\chi\,r^{1/3}\,\Im\,e\;.
\end{equation}
In the limit $\chi\to0$, we see that $\deltabeta$, $\Re\,e$ and $s$ converge to their value in the scaling $c=0$, while $\deltaalpha\simeq0.107018\,\chi^{2}$ and $\Im\,e\simeq-0.224251\,\chi/r$, with the same numerical constants as in the scaling $c=0$.\\\indent
The limit $\chi\to\infty$ is a bit more complicated. We check that $\deltaalpha\simeq\sqrt{3}\chi/12$ and $\deltabeta\simeq\sqrt{3}\chi/2$ in this limit. Indeed, with these values for $\deltaalpha$ and $\deltabeta$, we observe that $\Phi_{+}(\chi u)>0$ for all $u>0$, while $\Phi_{-}(\chi u)<0$ when $u<3/(4\pi)$ and $\Phi_{-}(\chi u)>0$ otherwise. In the integrals of (\ref{deltabeta c=1/3}) and (\ref{deltaalpha c=1/3}), the first term of the integrands gives a contribution exponentially small in $\chi$ to the integral, while the second term of the integrands contributes only for $v/\chi<3/(4\pi)$, with $\rme^{\Phi_{-}(v)}\to0$. Performing explicitly the integrals then shows that (\ref{deltabeta c=1/3}) and (\ref{deltaalpha c=1/3}) are indeed verified. We also checked numerically that the solutions of (\ref{deltabeta c=1/3}) and (\ref{deltaalpha c=1/3}) for finite $\chi$ seem to grow as $\deltaalpha\simeq\sqrt{3}\chi/12$ and $\deltabeta\simeq\sqrt{3}\chi/2$ for large $\chi$.\\\indent
Similar calculations lead to $\Re\,e\simeq-2^{1/2}3^{3/4}\chi^{5/2}/(5\pi r^{5/3})$ and $\Im\,e\simeq-2^{1/2}\chi^{3/2}/(3^{1/4}\pi r)$ in the limit $\chi\to\infty$. Using (\ref{s c=1/3}), it implies $s=0$ on the scale $\chi^{5/2}$. Going beyond that requires some more work: one has to take into account the contribution of the integrals near $v=3\chi/(4\pi)$, making the change of variables $v=3\chi/(4\pi)+\chi^{-1/2}u$. Writing $\deltaalpha=\sqrt{3}\chi/12+\chi^{-2}\deltaalpha_{2}$ and $\deltabeta=\sqrt{3}\chi/2+\chi^{-2}\deltabeta_{2}$, one finds at the end of the calculation $\deltaalpha_{2}\simeq-\pi^{2}/144$, $\deltabeta_{2}\to0$, $\Re\,e\simeq-2^{1/2}3^{3/4}\chi^{5/2}/(5\pi r^{5/3})-\pi/(2^{7/2}3^{3/4}\chi^{1/2}r^{5/3})$ and $\Im\,e\simeq-2^{1/2}\chi^{3/2}/(3^{1/4}\pi r)+\pi/(2^{7/2}3^{3/4}\chi^{3/2}r)$. It finally leads to $s\simeq\pi/(2^{1/2}3^{3/4}\chi^{1/2}r^{2/3})$.
\end{subsection}

\begin{subsection}{Scalings \texorpdfstring{$c<0$}{c<0} and \texorpdfstring{$0<c<1/3$}{0<c<1/3}.}
The crossover scaling $c=0$ is surrounded by the $2$ regimes $c<0$ and $0<c<1/3$. Similar calculations to the ones performed for the scalings $c=0$ allow to compute the quantities $\alpha$, $\beta$, $\Re\,e$, $\Im\,e$ and $s$ at leading order in $r$.\\\indent
We observe that the results found in the regime $c<0$ are identical to the limit $\chi\to0$ in the scaling $c=0$. Similarly, the results found in the regime $0<c<1/3$ are identical to the limit $\chi\to\infty$ in the scaling $c=0$.\\
Since the regimes $c<0$ and $0<c<1/3$ differ only by the value of the parameter $\deltaalpha$, which is just an intermediate quantity needed for the calculations, one can for all purpose consider this as a unique regime, $c<1/3$: the quantities of interest $e$ and $s$ then depend in a simple way on $r$, $\chi$ and $c$ in the whole regime.
\end{subsection}

\begin{subsection}{Scaling \texorpdfstring{$1/3<c<1$}{1/3<c<1}.}
The regime $1/3<c<1$ is much more complicated. There, one is lead to make a change of variable of the form $u=\pm1/4+d_{1}/r^{1-c}+d_{2}/r^{2(1-c)}+\ldots+d_{m}/r^{m(1-c)}+v/r^{(c+1)/2}$ in the integrals, where the constants $d_{j}$ must be such that the argument of the exponential in $\sigma(u)$ is of order $r^{0}$. Using a similar rewriting to (\ref{intF}), but with $\pm1/4$ replaced by $\pm1/4+d_{1}/r^{1-c}+d_{2}/r^{2(1-c)}+\ldots+d_{m}/r^{m(1-c)}$, the same kind of calculations as in the other scalings can in principle be done.\\\indent
We checked only the specific case $c=2/3$. There, making the change of variables $u=\pm1/4+d_{1}/r^{1/3}+d_{2}/r^{2/3}+v/r^{5/6}$ in the integrals, we find that $d_{1}$ must be solution of the equation
\begin{equation}
(3\chi-4\pi d_{1})\Re[(\deltabeta+2\rmi\pi d_{1})^{1/2}]=2(\deltabeta-6\deltaalpha)\Im[(\deltabeta+2\rmi\pi d_{1})^{1/2}]\;,
\end{equation}
while $d_{2}$ has a rather complicated (but completely explicit) expression in terms of $\chi$, $\deltaalpha$, $\deltabeta$ and $d_{1}$. The equation for $\beta$ at order $r^{-1/2}$ then gives
\begin{equation}
\Re[(\deltabeta+2\rmi\pi d_{1})^{3/2}]=0\;,
\end{equation}
while the equation for $\alpha$ at order $r^{1/2}$ leads to
\begin{equation}
3\chi-4\pi d_{1}=2\rmi(\deltabeta-6\deltaalpha)\;.
\end{equation}
Gathering the last $3$ equations, one finds $d_{1}=3\chi/(4\pi)$, $\deltaalpha=\sqrt{3}\chi/12$ and $\deltabeta=\sqrt{3}\chi/2$. The equation for $d_{2}$ then gives $d_{2}=-7\chi^{2}/(10\sqrt{3}\pi)$. From the equation for $e$, one obtains $\Re\,e=-2^{1/2}3^{3/4}\chi^{5/2}/(5\pi r^{5/6})$ and $\Im\,e=-2^{1/2}\chi^{3/2}/(3^{1/4}\pi r^{1/2})$, while the equation for $s$ leads to $s=\pi/(2^{1/2}3^{3/4}\chi^{1/2}r^{5/6})$. We observe that the numerical constants are the same as in the limit $\chi\to\infty$ of the scaling $c=1/3$. We conjecture that this is the case for all the scaling $1/3<c<1$.
\end{subsection}
\end{section}

\begin{section}{Explicit calculations for the generating function \texorpdfstring{$f(t)$}{f(t)}}
\label{appendix trace}
In this appendix, we calculate explicitly the contour integrals in (\ref{rho[sigma]}), (\ref{e[sigma] tr}) and  (\ref{s[sigma]}), with $\sigma(z)$ given by (\ref{sigma(z) trace}). In order to do this, we first prove two useful formulas, (\ref{exp(psi)}) and (\ref{exp(phi)}) for the exponential of the functions $\psi$ (\ref{psi(z)}) and $\varphi$ (\ref{phi(z)}).

\begin{subsection}{Formula for \texorpdfstring{$\rme^{x\psi(z)}$}{exp(x psi(z))}}
From (\ref{phipsi[g]}), one has $\rme^{x\,\psi(z)}=(g^{-1}(z))^{x}$. Using (\ref{h(g-1)}) with $h(y)=y^{x}$, the expansion near $z=0$ leads to
\begin{equation}
\label{exp(psi)}
\rme^{x\,\psi(z)}=x\,\sum_{r=0}^{\infty}\C{\rho\,r+x}{r}\frac{(-1)^{r}z^{r}}{\rho\,r+x}\;.
\end{equation}
\end{subsection}

\begin{subsection}{Formula for \texorpdfstring{$\rme^{-x\varphi(z)}$}{exp(-x phi(z))}}
Expanding the exponential in $\rme^{-x\varphi(z)}$ and using (\ref{relation phi psi 2}), one finds
\begin{equation}
\rme^{-x\varphi(z)}=1+\sum_{k=1}^{\infty}\frac{(-z)^{-k}x^{k}}{k!}\rme^{k(1-\rho)\psi(z)}\;.
\end{equation}
Eq.~(\ref{exp(psi)}) then leads to
\begin{equation}
\label{exp(phi)}
\hspace{-5mm}
\rme^{-x\varphi(z)}=1+(1-\rho)\sum_{k=1}^{\infty}\frac{x^{k}}{(k-1)!}\sum_{r=0}^{\infty}\C{\rho\,r+(1-\rho)k}{r}\frac{(-1)^{r-k}z^{r-k}}{\rho\,r+(1-\rho)k}\;.
\end{equation}
In particular, one has
\begin{equation}
\label{residue(exp(phi))}
\oint\frac{\rmd z}{2\rmi\pi}\,\frac{\rme^{-x\varphi(z)}}{z}=\rho+(1-\rho)\rme^{x}\;.
\end{equation}
\end{subsection}

\begin{subsection}{Exact expression for the Lagrange multiplier \texorpdfstring{$\lambda$}{lambda}}
The Lagrange multiplier $\lambda$ is fixed by the normalization condition (\ref{rho[sigma]}) with $\sigma$ given by (\ref{sigma(z) trace}). One has
\begin{equation}
\rho=\sum_{r=0}^{\infty}(-1)^{r}(\rme^{-\lambda})^{r}\oint\frac{\rmd z}{2\rmi\pi}\,\frac{\rme^{-rt\varphi(z)}}{z}\;.
\end{equation}
Using (\ref{residue(exp(phi))}) with $x=rt$ implies
\begin{equation}
\rho=\frac{\rho}{1+\rme^{-\lambda}}+\frac{1-\rho}{1+\rme^{t-\lambda}}\;.
\end{equation}
Solving for $\lambda$ finally gives
\begin{equation}
\label{lambda(t,rho)}
\lambda=\log\frac{2\rho\,\rme^{t}}{1-2\rho+\sqrt{1+4\rho(1-\rho)(\rme^{t}-1)}}\;,
\end{equation}
which simplifies at half filling to $\lambda=t/2$. The sign $+$ is chosen in front of the square root for continuity at $t=0$, for which one has $\lambda=\log[\rho/(1-\rho)]$.
\end{subsection}

\begin{subsection}{Exact expression for the eigenvalue \texorpdfstring{$e$}{e}}
The expression (\ref{e[sigma] tr}) for the eigenvalue can be made completely explicit. From (\ref{sigma(z) trace}), one has
\begin{eqnarray}
&& e=\sum_{r=0}^{\infty}(-1)^{r}(\rme^{-\lambda})^{r}\oint\frac{\rmd z}{2\rmi\pi}\,\frac{\rme^{-rt\varphi(z)}\varphi(z)}{z}\nonumber\\
&&\hspace{2mm} =\oint\frac{\rmd z}{2\rmi\pi}\,\frac{\varphi(z)}{z}+\partial_{t}\Big[\sum_{r=1}^{\infty}\frac{(-1)^{r-1}(\rme^{-\lambda})^{r}}{r}\oint\frac{\rmd z}{2\rmi\pi}\,\frac{\rme^{-rt\varphi(z)}}{z}\Big]\;.
\end{eqnarray}
Using (\ref{residue(exp(phi))}) to compute the residue, one finds
\begin{equation}
e=-\frac{1-\rho}{1+\rme^{t-\lambda}}=\frac{1-\sqrt{1+4\rho(1-\rho)(\rme^{t}-1)}}{2(\rme^{t}-1)}\;.
\end{equation}
\end{subsection}

\begin{subsection}{Exact expression for \texorpdfstring{$s$}{s}}
After a little rewriting, the expression (\ref{s[sigma]}) for $s$ becomes
\begin{equation}
s=\oint\frac{\rmd z}{2\rmi\pi z}\,\Big[(1-\sigma(z))\big(\lambda+t\varphi(z)\big)+\log\big(1+\rme^{-\lambda-t\varphi(z)}\big)\Big]\;.
\end{equation}
Using (\ref{rho[sigma]}), (\ref{e[sigma] tr}) and (\ref{sigma(z) trace}), one has
\begin{equation}
s=\lambda(1-\rho)-t(1-\rho)-te+\sum_{r=1}^{\infty}\frac{(-1)^{r-1}\rme^{r\lambda}}{r}\oint\frac{\rmd z}{2\rmi\pi}\,\frac{\rme^{-rt\varphi(z)}}{z}\;.
\end{equation}
Using (\ref{residue(exp(phi))}) to compute the residue, one finds
\begin{equation}
s=-te+\rho\log(1+\rme^{-\lambda})+(1-\rho)\log(1+\rme^{\lambda-t})\;.
\end{equation}
\end{subsection}
\end{section}

\begin{section}{Free particles}
\label{appendix free particles}
In this appendix, we study a system of $N=\rho L$ non-interacting particles hopping to the nearest site on the right with rate $1$ on a periodic lattice of $L$ sites. Unlike TASEP, there is no exclusion constraint. We will consider both the case of distinguishable particles and the case of undistinguishable particles.\\\indent
Similarly to TASEP, we call $M(\gamma)$ the deformation of the Markov matrix which counts the current of particles. The action of $M(\gamma)$ on a microstate with particles at positions $x_{1},\ldots,x_{N}$ is
\begin{equation}
M(\gamma)|x_{1},\ldots,x_{N}\rangle=\sum_{j=1}^{N}\big(\rme^{\gamma}|\ldots,x_{j}+1,\ldots\rangle-|\ldots,x_{j},\ldots\rangle\big)\;.
\end{equation}
The $x_{j}$'s need not be distinct. For distinguishable particles, the $j$-th element of $|x_{1},\ldots,x_{N}\rangle$ is the position of the $j$-th particle, and the total number of microstates is $|\Omega_{\text{free}}^{\text{d}}|=L^{N}$. For undistinguishable particles, the positions are kept ordered as $x_{1}\leq\ldots\leq x_{N}$, and the number of configurations is then $|\Omega_{\text{free}}^{\text{u}}|=\C{L+N-1}{N}$. In both cases, the eigenvectors of $M(\gamma)$ are products of plane waves, and the eigenvalues are of the form
\begin{equation}
\label{E free}
E=\sum_{j=1}^{N}\big(\rme^{\gamma-2\rmi\pi k_{j}/L}-1\big)\;,
\end{equation}
where each $k_{j}$ is an integer between $1$ and $L$. For distinguishable particles, there is no further restriction on the $k_{j}$'s. For undistinguishable particles the $k_{j}$'s must be ordered, $k_{1}\leq\ldots\leq k_{N}$.\\\indent
As in the case of TASEP, we define a density of eigenvalues $D(e)$ as in (\ref{D[path integral]}) and a quantity $f(t)$ as in (\ref{f[M]}). We calculate in this appendix $D(e)$ in the thermodynamic limit in the case of undistinguishable particles, and $f(t)$ for both distinguishable and undistinguishable particles.

\begin{subsection}{Density of eigenvalues for undistinguishable particles}
\begin{figure}
  \begin{center}
    \begin{tabular}{ccc}
      \begin{tabular}{c}\includegraphics[width=70mm]{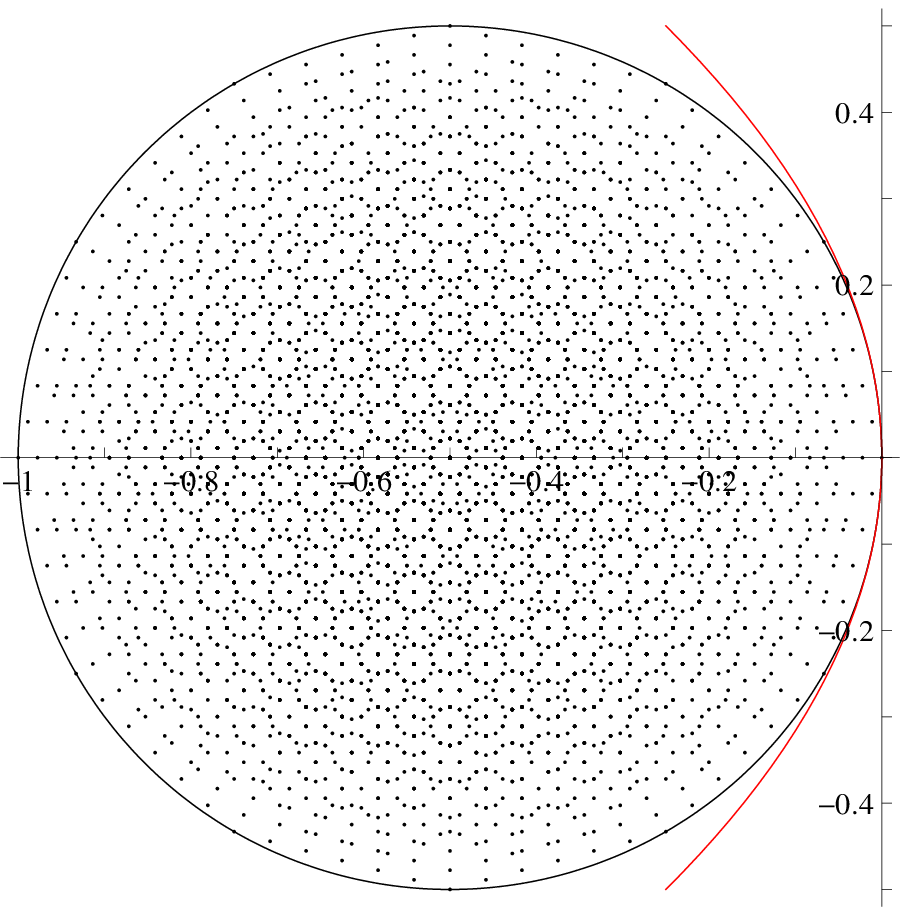}\end{tabular}
      &&
      \begin{tabular}{c}\includegraphics[width=70mm]{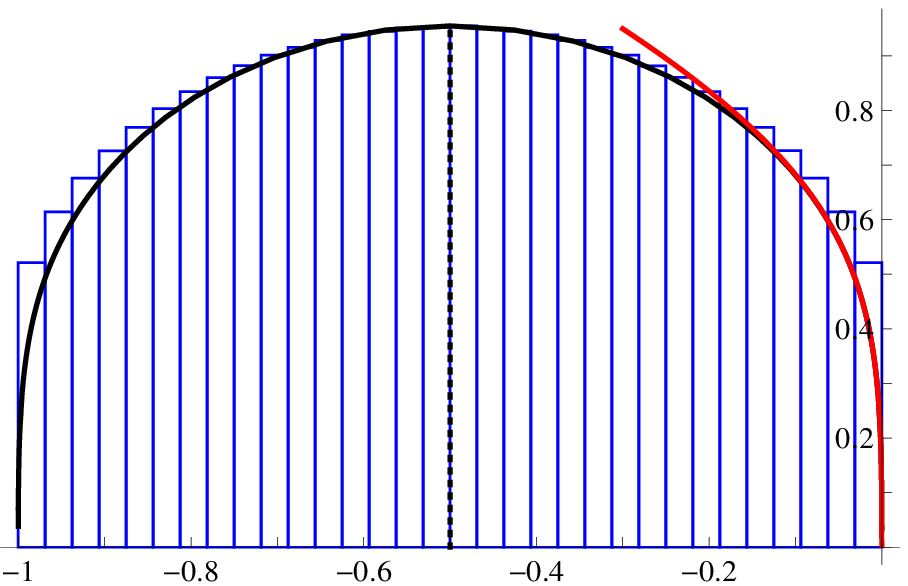}\end{tabular}
    \end{tabular}
  \end{center}
  \caption{On the left, spectrum of the Markov matrix $M$ for a system of $N=6$ non-interacting undistinguishable particles on $L=12$ sites. The black dots are the eigenvalues rescaled by a factor $1/L$. The black circle is the edge of the spectrum in the thermodynamic limit with $\rho=1/2$. The red (gray in printed version) curve is the parabolic approximation of the circle near the origin. On the right, number of rescaled eigenvalues with a given real part $e$ for non-interacting undistinguishable particles at half-filling, as a function of $e$. The black curve corresponds to the expressions (\ref{s[eta] free U}), (\ref{e[eta] free U}) parametrized by $\omega$ ranging from $-1000$ to $1000$. The red (gray in printed version) curve is the asymptotics (\ref{s(e) free U}). The histograms correspond to the density of real part of eigenvalues obtained from numerical diagonalization for the finite system with $N=9$ particles on $L=18$ sites. The histograms are shifted so that their maximum is $(\log2+3\log3)/2\simeq L^{-1}\log|\Omega_{\text{free}}^{\text{u}}|$, with $|\Omega_{\text{free}}^{\text{u}}|$ the total number of microstates.}
  \label{fig s(e) free U}
\end{figure}
Defining the density profile $\eta(u)$ of the $k_{j}$'s as in section \ref{subsection eigenvalues thermodynamic limit}, the rescaled eigenvalue $e=E/L$ is expressed in terms of $\eta$ as
\begin{equation}
\label{e[eta] free U}
e[\eta]=\int_{0}^{1}\rmd u\,\eta(u)(\rme^{\gamma-2\rmi\pi u}-1)\;.
\end{equation}
The number $\Omega[\eta]$ of ways to place $L\,\eta(u)\,du$ $k_{j}$'s in an interval of length $L\,du$ is $\C{L\,du+L\,\eta(u)\,du-1}{L\,\eta(u)\,du}$. Stirling's formula implies $\Omega[\eta]\sim\rme^{Ls}$, where the "entropy" $s[\eta]$ is
\begin{equation}
\label{s[eta] free U}
s[\eta]=\int_{0}^{1}\rmd u\,[-\eta(u)\log\eta(u)+(1+\eta(u))\log(1+\eta(u))]\;.
\end{equation}
The difference with eq.~(\ref{s[eta]}) for TASEP comes from the fact that several $k_{j}$'s can be equal for non-interacting particles.\\\indent
We introduce the two Lagrange multipliers $\lambda\in\mathbb{R}$ and $\omega\in\mathbb{C}$ as in (\ref{s + Lagrange multipliers}). The optimal function $\eta^{*}(u)$ that maximizes (\ref{s + Lagrange multipliers}) with $s[\eta]$ given by (\ref{s[eta] free U}) is
\begin{equation}
\label{eta* free U}
\eta^{*}(u)=\Big(-1+\rme^{-\lambda-2\Re[\omega(\rme^{\gamma-2\rmi\pi u}-1)]}\Big)^{-1}\;.
\end{equation}
Solving numerically (\ref{rho[eta]}) for several values of $\omega$ allows to plot $s(e)$, see fig.~\ref{fig s(e) free U}. We are interested in the limit $e\to0$, which corresponds to $\omega\to\infty$, $\gamma=0$. We will only treat the case $e\in\mathbb{R}$, for which $\omega\in\mathbb{R}$. We first expand the denominator and the exponential in the expression (\ref{eta* density}) of $\eta^{*}$, as
\begin{equation}
\eta^{*}(u)=\sum_{k=1}^{\infty}\sum_{j=0}^{\infty}\frac{k^{j}}{j!}\rme^{k(\lambda-2\omega)}(\omega\rme^{-2\rmi\pi u}+\overline{\omega}\rme^{2\rmi\pi u})^{j}\;,
\end{equation}
where $\overline{\,\cdot\,}$ denotes complex conjugation. The integral over $u$ can then be performed in (\ref{rho[eta]}), (\ref{e[eta] free U}) and (\ref{s[eta] free U}). After summing over $j$, we find
\begin{equation}
\rho=\sum_{k=1}^{\infty}\rme^{k(\lambda-2\omega)}\I_{0}(2k\omega)\;,
\end{equation}
\begin{equation}
e=-\rho+\sum_{k=1}^{\infty}\rme^{k(\lambda-2\omega)}\I_{1}(2k\omega)\;,
\end{equation}
and
\begin{equation}
s=-\rho\lambda-2\omega e+\sum_{k=1}^{\infty}\frac{\rme^{k(\lambda-2\omega)}}{k}\,\I_{0}(2k\omega)\;.
\end{equation}
In the previous expressions, $\I_{0}$ and $\I_{1}$ are modified Bessel functions of the first kind. Taking the asymptotics of the Bessel functions for large argument and summing over $j$ gives
\begin{equation}
\sqrt{2\pi}\rho\simeq\frac{\Li_{1/2}(\rme^{\lambda})}{\sqrt{2\omega}}+\frac{\Li_{3/2}(\rme^{\lambda})}{8(2\omega)^{3/2}}+\frac{9\,\Li_{5/2}(\rme^{\lambda})}{128(2\omega)^{5/2}}\;,
\end{equation}
\begin{equation}
\sqrt{2\pi}(e+\rho)\simeq\frac{\Li_{1/2}(\rme^{\lambda})}{\sqrt{2\omega}}-\frac{3\Li_{3/2}(\rme^{\lambda})}{8(2\omega)^{3/2}}-\frac{15\,\Li_{5/2}(\rme^{\lambda})}{128(2\omega)^{5/2}}\;,
\end{equation}
and
\begin{equation}
\sqrt{2\pi}(s+\rho\lambda+2\omega e)\simeq\frac{\Li_{3/2}(\rme^{\lambda})}{\sqrt{2\omega}}+\frac{\Li_{5/2}(\rme^{\lambda})}{8(2\omega)^{3/2}}+\frac{9\,\Li_{7/2}(\rme^{\lambda})}{128(2\omega)^{5/2}}\;.
\end{equation}
In the limit $\omega\to\infty$, one has $\lambda\to0^{-}$. After expanding the polylogarithms, we finally obtain
\begin{eqnarray}
\label{s(e) free U}
&& s(e)\simeq\frac{3\,\zeta(3/2)^{2/3}(-e)^{1/3}}{2\pi^{1/3}}-\frac{\pi^{1/3}(-e)^{2/3}}{\rho\,\zeta(3/2)^{2/3}}\nonumber\\
&&\hspace{15mm} +\Big(\frac{2\pi}{3\rho^{2}\zeta(3/2)^{2}}+\frac{\zeta(1/2)}{\rho^{2}\zeta(3/2)}-\frac{\zeta(5/2)}{4\,\zeta(3/2)}\Big)e\;,
\end{eqnarray}
where $\zeta$ is Riemann zeta function. We observe that $s(e)$ vanishes for $e=0$ with an exponent $1/3$. This exponent should not be confused with the exponent $1/2$ obtained from $\Re\,E\simeq-2\pi^{2}\sum_{j=1}^{N}k_{j}^{2}/L^{2}$ for eigenvalues that do not scale proportionally with $L$.
\end{subsection}

\begin{subsection}{Function \texorpdfstring{$f(t)$}{f(t)} for distinguishable particles}
\begin{figure}
  \begin{center}
    \begin{tabular}{ccc}
      \begin{tabular}{c}\includegraphics[width=70mm]{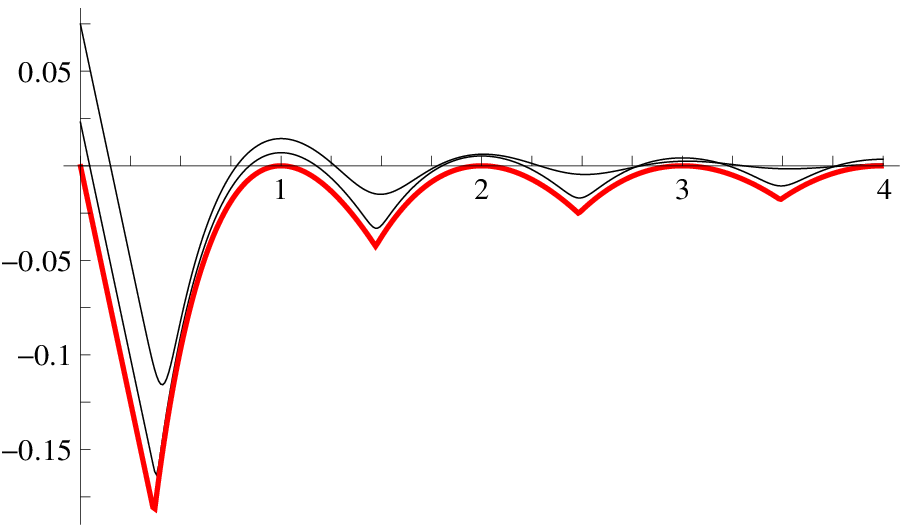}\end{tabular}
      &&
      \begin{tabular}{c}\includegraphics[width=70mm]{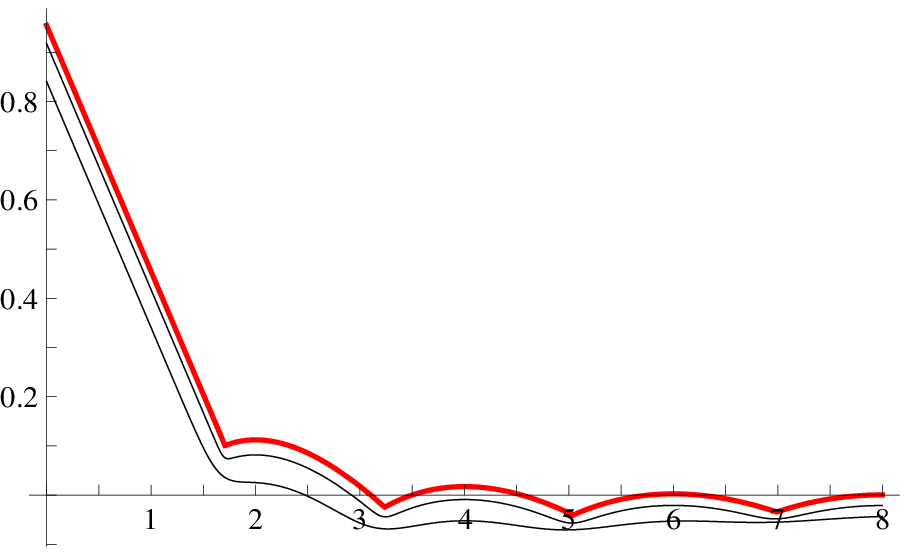}\end{tabular}
    \end{tabular}
  \end{center}
  \caption{Logarithm of the trace of the time evolution operator for non-interacting particles. On the left, graph of $f(\tau L)/L$ as a function of $\tau$ for distinguishable particles, with $f$ defined in (\ref{f[M]}). In black are exact computations for $\rho=1/2$, $\gamma=0$ with $L=20$ (upper curve) and $L=100$ (lower curve). The thick, red curve corresponds to the large $L$ limit (\ref{f free d asymptotic L}). On the right, graph of $f(t)$ as a function of $t$ for undistinguishable particles, with $f$ defined in (\ref{f[M]}). In black are exact computations for $\rho=1/2$, $\gamma=0$ with $L=20$ (lower curve) and $L=80$ (upper curve). The thick, red curve corresponds to the large $L$ limit (\ref{f free u asymptotic L}).}
  \label{fig f free}
\end{figure}
From the definition (\ref{f[M]}) and the expression (\ref{E free}) for the eigenvalues, one has
\begin{equation}
\label{f free d}
f(t)=\rho\log\Big(\sum_{k=1}^{L}\rme^{t(\rme^{\gamma-2\rmi\pi k/L}-1)}\Big)\;.
\end{equation}
For finite times, the sum becomes an integral in the large $L$ limit. After a rewriting as a contour integral, one finds
\begin{equation}
f(t)-\rho\log L=\rho\log\Big(\oint\frac{\rmd z}{z}\,\rme^{t(\rme^{\gamma}z-1)}\Big)=-\rho\,t\;.
\end{equation}
We are also interested in $f(t)$ for times $t$ of order $L$. Expanding the exponential in (\ref{f free d}) leads to
\begin{equation}
f(\tau L)=\rho\log\Big(\sum_{k=1}^{L}\sum_{j=0}^{\infty}\sum_{m=0}^{j}\C{j}{m}\frac{(-1)^{j-m}L^{j}\tau^{j}\rme^{m\gamma}\rme^{-2\rmi\pi km/L}}{j!}\Big)\;.
\end{equation}
Exchanging the order of the summations over $j$ and $m$ allows to perform the summation over $j$. One finds
\begin{equation}
f(\tau L)=\rho\log\Big(\sum_{k=1}^{L}\sum_{m=0}^{\infty}\frac{L^{m}\tau^{m}\rme^{-\tau L}\rme^{m\gamma}\rme^{-2\rmi\pi km/L}}{m!}\Big)\;.
\end{equation}
The summation over $k$ is then done with the help of
\begin{equation}
\label{sum k}
\sum_{k=1}^{L}\rme^{-2\rmi\pi km/L}=L\sum_{r=0}^{\infty}\delta_{m,rL}\;,
\end{equation}
which leads to
\begin{equation}
f(\tau L)-\rho\log L=\rho\log\Big(\sum_{r=0}^{\infty}\frac{L^{rL}\tau^{rL}\rme^{-\tau L}\rme^{r\gamma L}}{(rL)!}\Big)\;.
\end{equation}
Using Stirling's formula for $(rL)!$ (except for the term $r=0$) and extracting the leading term of the sum, one has
\begin{eqnarray}
\label{f free asymptotic L + corrections}
&& \frac{f(\tau L)}{L}\simeq\max_{r\in\mathbb{N}}\rho(-\tau+r(1+\log(\rme^{\gamma}\tau/r)))\nonumber\\
&&\hspace{20mm} -\openone_{\{r(\tau)\geq1\}}\frac{\rho\log(2\pi r(\tau)L)}{2L}-\openone_{\{r(\tau)\geq1\}}\frac{1}{12r(\tau)L^{2}}\;,
\end{eqnarray}
with the convention $r\log r=0$ for $r=0$. In the second and third terms, $r(\tau)$ is the $r$ corresponding to the maximum in the first term. Defining $\tau_{0}=0$ and for $r\in\mathbb{N}^{*}$
\begin{equation}
\rme^{\gamma}\tau_{r}=\frac{\rme^{-1}r^{r}}{(r-1)^{r-1}}\;,
\end{equation}
one finally finds for $\tau_{r}\leq\tau\leq\tau_{r+1}$
\begin{equation}
\label{f free d asymptotic L}
\lim_{L\to\infty}\frac{f(\tau L)}{L}=-\rho\,\tau+\rho\,r+\rho\,r\log\frac{\rme^{\gamma}\tau}{r}\;.
\end{equation}
For large $r$, one has $\rme^{\gamma}\tau_{r}\simeq r-1/2$, hence for large $\tau$
\begin{equation}
\lim_{L\to\infty}\frac{f(\tau L)}{L}\simeq\rho(\rme^{\gamma}-1)\tau-\frac{\rho(\rme^{\gamma}\tau-[\rme^{\gamma}\tau])^{2}}{2\rme^{\gamma}\tau}\;,
\end{equation}
where $[\rme^{\gamma}\tau]$ is the integer closest to $\rme^{\gamma}\tau$.
\end{subsection}

\begin{subsection}{Function \texorpdfstring{$f(t)$}{f(t)} for undistinguishable particles}
From the definition (\ref{f[M]}) and the expression (\ref{E free}) for the eigenvalues, one has
\begin{equation}
f(t)=\frac{1}{L}\log\Big(\sum_{1\leq k_{1}\leq\ldots\leq k_{N}\leq L}\prod_{i=1}^{N}\rme^{t(\rme^{\gamma-2\rmi\pi k_{i}/L}-1)}\Big)\;.
\end{equation}
Expanding the exponential leads to
\begin{equation}
\fl\hspace{5mm}
f(t)=\frac{1}{L}\log\Big(\sum_{1\leq k_{1}\leq\ldots\leq k_{N}\leq L}\prod_{i=1}^{N}\sum_{j=0}^{\infty}\sum_{m=0}^{j}\C{j}{m}\frac{(-1)^{j-m}t^{j}\rme^{m\gamma}\rme^{-2\rmi\pi k_{i}m/L}}{j!}\Big)\;.
\end{equation}
Exchanging the order of the summations over $j$ and $m$ allows to perform the summation over $j$. One finds
\begin{equation}
f(t)=\frac{1}{L}\log\Big(\sum_{1\leq k_{1}\leq\ldots\leq k_{N}\leq L}\prod_{i=1}^{N}\sum_{m=0}^{\infty}\frac{t^{m}\rme^{-t}\rme^{m\gamma}\rme^{-2\rmi\pi k_{i}m/L}}{m!}\Big)\;.
\end{equation}
So far, the calculation parallels the one for distinguishable particles in the scaling $t\sim L$. In order to perform the summation over the $k_{i}$, we first use the relation
\begin{equation}
\label{S(N,L)}
\hspace{-5mm}
S_{N}(L)=\!\!\!\sum_{1\leq k_{1}\leq\ldots\leq k_{N}\leq L}\prod_{i=1}^{N}f(k_{i})=\oint\frac{\rmd z}{2\rmi\pi z^{N+1}}\exp\Big[\sum_{a=1}^{\infty}\sum_{k=1}^{L}\frac{z^{a}f(k)^{a}}{a}\Big]\;,
\end{equation}
The contour integral is over a contour enclosing $0$. Eq.~(\ref{S(N,L)}) can be proved by considering the formal series in $z$
\begin{equation}
\sum_{N=0}^{\infty}z^{N}S_{N}(L)=\prod_{k=1}^{L}\frac{1}{1-z f(k)}\;.
\end{equation}
Eq.~(\ref{S(N,L)}) gives
\begin{equation}
\hspace{-20mm}
f(t)=\frac{1}{L}\log\Big(\oint\frac{\rmd z}{2\rmi\pi}\frac{\rme^{-Nt}}{z^{N+1}}\exp\Big[\sum_{a=1}^{\infty}\frac{z^{a}}{a}\sum_{k=1}^{L}\sum_{m_{1},\ldots,m_{a}=0}^{\infty}\prod_{i=1}^{a}\frac{t^{m_{i}}\rme^{m_{i}\gamma}\rme^{-2\rmi\pi km_{i}/L}}{m_{i}!}\Big]\Big)\;.
\end{equation}
Using (\ref{sum k}) to sum over $k$ leads to
\begin{equation}
\hspace{-20mm}
f(t)=\frac{1}{L}\log\Big(\oint\frac{\rmd z}{2\rmi\pi}\frac{\rme^{-Nt}}{z^{N+1}}\exp\Big[L\sum_{a=1}^{\infty}\frac{z^{a}}{a}\sum_{r=0}^{\infty}\sum_{m_{1},\ldots,m_{a}=0}^{\infty}\delta_{rL,\sum\limits_{i=1}^{a}m_{i}}\prod_{i=1}^{a}\frac{t^{m_{i}}\rme^{m_{i}\gamma}}{m_{i}!}\Big]\Big)\;.
\end{equation}
The multinomial sum over the $m_{i}$'s can be performed. One has
\begin{equation}
f(t)=\frac{1}{L}\log\Big(\oint\frac{\rmd z}{2\rmi\pi}\frac{\rme^{-Nt}}{z^{N+1}}\exp\Big[L\sum_{a=1}^{\infty}\frac{z^{a}}{a}\sum_{r=0}^{\infty}\frac{(a\,\rme^{\gamma}t)^{rL}}{(rL)!}\Big]\Big)\;.
\end{equation}
The summation over $a$ can be done explicitly. For $r\geq1$, it gives a polylogarithm. One finds
\begin{equation}
f(t)=\frac{1}{L}\log\Big(\oint\frac{\rmd z}{2\rmi\pi}\frac{\rme^{-Nt}}{z^{N+1}(1-z)^{L}}\exp\Big[L\sum_{r=1}^{\infty}\frac{(\rme^{\gamma}t)^{rL}}{(rL)!}\Li_{1-rL}(z)\Big]\Big)\;.
\end{equation}
We deform the contour of integration so that it encloses the negative real axis. In the thermodynamic limit, it is then possible to use the asymptotics
\begin{equation}
\Li_{-n}(z)\simeq\Gamma(n+1)(-\log z)^{-n-1}\;
\end{equation}
for large $n$. It leads to
\begin{equation}
f(t)\simeq\frac{1}{L}\log\Big(\oint\frac{\rmd z}{2\rmi\pi}\frac{\rme^{-Nt}}{z^{N+1}(1-z)^{L}}\exp\Big[\sum_{r=1}^{\infty}\frac{1}{r}\Big(-\frac{\rme^{\gamma}t}{\log z}\Big)^{rL}\Big]\Big)\;.
\end{equation}
Summing explicitly over $r$ gives
\begin{equation}
f(t)\simeq\frac{1}{L}\log\Big(\oint\frac{\rmd z}{2\rmi\pi}\frac{\rme^{-Nt}}{z^{N+1}(1-z)^{L}\Big(1-\Big(-\frac{\rme^{\gamma}t}{\log z}\Big)^{L}\Big)}\Big)\;.
\end{equation}
Expanding the last factor of the denominator, we finally obtain
\begin{equation}
f(t)\simeq\frac{1}{L}\log\Big(\sum_{r=0}^{\infty}\oint\frac{\rmd z}{2\rmi\pi}\frac{\rme^{-Nt}\Big(-\frac{\rme^{\gamma}t}{\log z}\Big)^{rL}}{z^{N+1}(1-z)^{L}}\Big)\;.
\end{equation}
The thermodynamic limit of $f(t)$ is extracted by calculating the saddle point $z_{r}$ of the contour integral. One finds
\begin{equation}
\label{f free u asymptotic L}
f(t)\simeq\max_{r\in\mathbb{N}}\Big(-\rho\,t-\rho\log z_{r}-\log(1-z_{r})+r\log\Big(-\frac{\rme^{\gamma}t}{\log z_{r}}\Big)\Big)\;,
\end{equation}
where $z_{r}$ verifies the equation
\begin{equation}
\frac{r}{\log z_{r}}=\frac{z_{r}}{1-z_{r}}-\rho\;.
\end{equation}
One has $z_{0}=\rho/(1+\rho)$. For $\rho=1/2$, we find numerically $z_{1}\simeq0.085$, $z_{2}\simeq0.016$, $z_{3}\simeq0.0024$. We introduce times $t_{r}$, $r\in\mathbb{N}$ such that when $t_{r}<t<t_{r+1}$, the saddle point that dominates (\ref{f free u asymptotic L}) is $z_{r}$. The value of $t_{r}$ is determined by continuity of $f(t)$. One has $t_{0}=0$, and for $\rho=1/2$, $\rme^{\gamma}t_{1}=1.7085$, $\rme^{\gamma}t_{2}=3.236$, $\rme^{\gamma}t_{3}=5.040$.\\\indent
For large $r$, $z_{r}$ decreases to $0$ as $z_{r}\simeq\rme^{-r/\rho}$. This implies, for $t_{r}<t<t_{r+1}$,
\begin{equation}
f(t)\simeq-\rho t+r+r\log\frac{\rho\,\rme^{\gamma}t}{r}\;,
\end{equation}
with $t_{r}$ given for large $r$ by
\begin{equation}
\label{t(r) free u asymptotic t}
\rho\,\rme^{\gamma}t_{r}\simeq\frac{\rme^{-1}r^{r}}{(r-1)^{r-1}}\simeq r-\frac{1}{2}\;.
\end{equation}
For large times, we finally obtain
\begin{equation}
f(t)\simeq\rho(\rme^{\gamma}-1)t-\frac{(\rho\,\rme^{\gamma}t-[\rho\,\rme^{\gamma}t])^{2}}{2\rho\,\rme^{\gamma}t}\;,
\end{equation}
where $[\rho\,\rme^{\gamma}t]$ is the integer closest to $\rho\,\rme^{\gamma}t$. This expression is very similar to the one found for distinguishable particles on the scale $t\sim L$.
\end{subsection}
\end{section}

\section*{References}


\end{document}